\documentclass[conference]{IEEEtran}
\IEEEoverridecommandlockouts
    
\pdfoutput=1

\usepackage[utf8]{inputenc} % norme pour encodage
\usepackage[T1]{fontenc}    % accents
\usepackage{microtype}      % meilleure mise en page

\usepackage{xspace}
\usepackage{todonotes}
\usepackage{listings}
\usepackage{multirow}
\usepackage{xcolor}
\usepackage{ amssymb }
\usepackage{amsmath}
\usepackage{amsfonts}
\usepackage{textpos}
\usepackage{graphicx}

\usepackage{caption}    % to avoid issues with acm style
\usepackage{subcaption}
\usepackage[ruled,lined,linesnumbered]{algorithm2e} 

\usepackage{latexsym}
\usepackage{euscript, stmaryrd, bigstrut}
\usepackage{float}
\usepackage{url}
\floatstyle{boxed}
\restylefloat{figure}

\usepackage{rotating}

\newcommand{\myred}[1]{\textcolor{red}{#1}}
\xdefinecolor{mongris}{rgb}{0.5,0.5,0.5}
\xdefinecolor{monvert}{rgb}{0,.5, 0}
\xdefinecolor{Rouge}{rgb}{1,0,0}
\xdefinecolor{seborange}{rgb}{1,0.75,0}

\newcommand{\codisasm}{\textsc{Codisasm}\xspace}
\newcommand{\binsec}{\textsc{Binsec}\xspace}
\newcommand{\binsecse}{\textsc{Binsec/se}\xspace}
\newcommand{\pinsec}{{\sc Pinsec}\xspace}
\newcommand{\idasec}{{\sc Idasec}\xspace}
\newcommand{\ret}{\texttt{ret}\xspace}
\newcommand{\call}{\texttt{call}\xspace}

\newcommand{\alignedd}{\texttt{[aligned]}\xspace}
\newcommand{\disalignedd}{\texttt{[disaligned]}\xspace}
\newcommand{\single}{\texttt{[single]}\xspace}
\newcommand{\multiple}{\texttt{[multiple]}\xspace}
\newcommand{\genuine}{\texttt{[genuine]}\xspace}
\newcommand{\violated}{\texttt{[violated]}\xspace}

\newcommand{\rouge}[1]{\textcolor{red}{#1}}

\newcommand{\monvert}[1]{\textcolor{monvert}{#1}}

\newcommand{\mycommentAnonymous}[1]{}   % anonymiser
\newcommand{\mycomment}[1]{}
\newcommand{\myparagraph}[1]{\medskip \noindent{\bf #1.}}
\newcommand{\mypar}{\smallskip \indent}

\newcommand{\forjournal}[1]{}

\newcommand{\mycheckmark}{\monvert{\bf   \checkmark}}
\newcommand{\good}{\mycheckmark}

\newcommand{\mybadmark}{\rouge{\bf   $\times$}}
\newcommand{\bad}{\mybadmark}

\newcommand{\xtunnel}{\mbox{{\sc X-Tunnel}}\xspace}
\newcommand{\xagent}{\mbox{{\sc X-Agent}}\xspace}
\newcommand{\bbdse}{{\sc bb-dse}\xspace}
\newcommand{\aspack}{\texttt{ASPack}\xspace}
\newcommand{\ollvm}{O-LLVM\xspace}
\newcommand{\ZE}{{\tt Z3}\xspace}
\newcommand{\objdump}{{\tt Objdump}\xspace}
\newcommand{\IDA}{{\tt IDA}\xspace}

\newcommand{\unsat}{\textsc{unsat}\xspace}
\newcommand{\coreutils}{\texttt{coreutils}\xspace}

\newcommand{\sdag}{$^{\dagger}$}
\newcommand{\sddag}{$^{\ddagger}$}

\definecolor{lblue}{HTML}{2C57AF}
\definecolor{dkgreen}{rgb}{0,0.6,0}
\newcommand{\myblue}[1]{\textcolor{lblue}{#1}}
\newcommand{\mygreen}[1]{\textcolor{dkgreen}{#1}}
\newcommand{\myorange}[1]{\textcolor{orange}{#1}}

\lstdefinestyle{sourcecode}{
  numbers=left,
  numbersep=-8pt,
  belowcaptionskip=1\baselineskip,
  breaklines=true,
  xleftmargin=\parindent,
  language=C,
  showstringspaces=false,
  basicstyle=\footnotesize\ttfamily,
  keywordstyle=\bfseries\color{dkgreen},
  commentstyle=\itshape\color{gray},
  identifierstyle=\color{blue},
  stringstyle=\color{mauve},
  numberstyle=\ttfamily
}

\begin{document}

%\toappear{}   %% gere copyright and co pour camera ready et soumission

%% % Copyright
%% \setcopyright{acmcopyright}
%% %\setcopyright{acmlicensed}
%% %\setcopyright{rightsretained}
%% %\setcopyright{usgov}
%% %\setcopyright{usgovmixed}
%% %\setcopyright{cagov}
%% %\setcopyright{cagovmixed}

%% % DOI
%% \doi{}

%% % ISBN
%% \isbn{}

%% %Conference
%% \conferenceinfo{name}{lieu}

%% \acmPrice{}

%\date{}

%\title{Backward-Bounded DSE: Targeting infeasibility proofs on Obfuscated Codes\mycommentAnonymous{\titlenote{Work partially funded by ANR, grant 12-INSE-0002.}}}
%\title{Backward-Bounded DSE: \\ Targeting Infeasibility Questions Arising in Deobfuscation \mycommentAnonymous{\titlenote{Work partially funded by ANR, grant 12-INSE-0002.}}}
%\title{Solving Infeasibility Questions Arising in Deobfuscation \\ with Backward-Bounded DSE  \mycommentAnonymous{\titlenote{Work partially funded by ANR, grant 12-INSE-0002.}}}

%\title{Solving Infeasibility Questions Arising in Deobfuscation \mycommentAnonymous{\titlenote{Work partially funded by ANR, grant 12-INSE-0002.}}}
\title{Targeting Infeasibility Questions \\ on Obfuscated Codes$^{\star}$ {\thanks{\noindent $^{\star}$ Work partially funded by ANR, grant 12-INSE-0002.}}}

\author{
\IEEEauthorblockN{Robin David, S\'ebastien Bardin}
\IEEEauthorblockA{CEA, LIST\\         %, PC 174\\
91191 Gif-sur-Yvette, France\\
\textit{firstname.lastname@cea.fr}}
\and
\IEEEauthorblockN{Jean-Yves Marion}
\IEEEauthorblockA{Université de Lorraine, CNRS \\ LORIA, Nancy, France \\ \textit{firstname.lastname@loria.fr}  } 
}

\maketitle

% Use the following at camera-ready time to suppress page numbers.
% Comment it out when you first submit the paper for review.
%\thispagestyle{empty}

%\vspace{-0.5cm}

\begin{abstract}
Software deobfuscation is a crucial activity in security analysis and especially, in malware analysis.  
While standard static and dynamic approaches suffer from well-known shortcomings, % (either in terms of robustness or exhaustiveness),  
Dynamic Symbolic Execution (DSE) has recently been proposed has an interesting alternative, more robust than static analysis and more complete  
than  dynamic analysis. 
Yet, DSE addresses certain kinds of questions encountered by a reverser namely feasibility questions.  
Many issues arising during reverse, e.g.~detecting protection schemes such as opaque predicates % or call stack tampering, 
 fall into the category of infeasibility questions. %, for which DSE does not fit.    
% DSE peut si : tous les chemins (impossible), très longs chemins (scale !), no sous approximation (pas ça en pratique, scale ! robust !)
%
%
In this article, we present the Backward-Bounded DSE, a generic, precise, efficient and robust  method for solving infeasibility questions. % the disassembly of  obfuscated codes. %  disassembly.  
%
% 
%We demonstrate the benefit of the method for the identification of {\it opaque predicates} %(no false  positive, very few false negative) 
%and for the 
%classification of {\tt \small return} statements in case of {\it call stack tampering}. \myred{**SB: self-modif ??**}
%
We demonstrate the benefit of the method for opaque predicates and  %(no false  positive, very few false negative) 
call stack tampering, and give some insight for its usage for some other protection schemes. % {\it self-modification}.
Especially, the technique has successfully been used on state-of-the-art packers as well as   on the government-grade X-Tunnel malware -- allowing its entire deobfuscation.  
Backward-Bounded DSE does not supersede existing DSE approaches, but rather complements them by addressing infeasibility questions in a scalable and precise manner. 
%
%%%As an application we also shows how to leverage infeasibility-based obfuscation information on a government-grade malware
%%%\xtunnel thus allowing to de-obfuscate it entirely.
%
Following this line, we propose sparse disassembly, a combination of Backward-Bounded DSE and static disassembly able to { enlarge}  dynamic disassembly { in a guaranteed way}, hence 
getting the best of dynamic and static disassembly.
This work paves the way for robust, efficient and precise  disassembly tools for heavily-obfuscated binaries. 
%
%\todo[inline]{penser à relire}

%****** structure : 5 sentences:  context, challenge, idea of solution (twist), achievement (quickly), impact / consequence / how it changes the world

\end{abstract}

\IEEEpeerreviewmaketitle

%\todo[inline]{SB: se mettre d'accord sur safe/correct, complete/ almost complete/ quasi-complete}

\section{Introduction} \label{sec:intro}

\myparagraph{Context} 
Obfuscation \cite{surreptitious}  is a prevalent practice aiming at
protecting some functionalities or properties of a program. Yet, while its
legitimate goal is intellectual property protection, obfuscation is widely used
for malicious purposes. Therefore, (binary-level) software deobfuscation
is a crucial task in reverse-engineering, especially for malware analysis.

A first step of deobfuscation is to recover the most accurate
control-flow graph of the program ({\it disassembly}), i.e.~to recover all
instructions and branches of the program under analysis. This is already challenging
for non-obfuscated codes due to tricky (but common) low-level constructs~\cite{binary_not_easy}
like indirect control flow (computed jumps, {\small \tt jmp eax}) or the interleaving
of code and data. But the situation gets largely worst in the case of obfuscated codes. 

Standard disassembly  approaches are essentially divided into {\it static methods}
and {\it dynamic methods}. On one hand, static (syntactic) disassembly tools
such as \IDA or \objdump have the potential to cover the whole program. Nonetheless,
they are easily fooled by obfuscations such as code overlapping \cite{codisasm},
opaque predicates \cite{Collberg1998}, opaque constants~\cite{kruegel}, call stack
tampering~\cite{lakhotia_call} and self-modification \cite{miller-callstack}.  
On the other hand, dynamic analysis cover only  a few executions of the program
and might miss both  significant parts of the code and crucial behaviors.
{\it Dynamic Symbolic Execution} (DSE) \cite{GLM12,CS13} (a.k.a {\it concolic execution}) is a recent and fruitful
formal approach to automatic testing, has recently been proposed has an interesting
approach for disassembly \cite{ccs2015,sp15,loop,BardinHV11,BHKLNPS2007}, more {\it
robust} than static analysis and covering more instructions than dynamic analysis.  
{\it Currently, only dynamic analysis and DSE are robust enough to address heavily
	obfuscated codes.}

\myparagraph{Problem} Yet, these dynamic methods only address reachability issues, 
namely {\it feasibility questions}, i.e.~verifiying that certain events or setting can occur,
e.g.~that an instruction in the code is indeed reachable. 
Contrariwise, many questions encountered\ during reversing tasks are {\it infeasibility
questions}, i.e.~checking that certain events or settings cannot occur. It can be used either for
detecting obfuscation schemes, e.g.~detecting that a branch is dead (i.e. it cannot be taken) or to prove their absence,  
e.g.~proving that a computed jump cannot lead to an improper address.

These {\it infeasibility issues are currently a blind spot of both standard and
advanced disassembly methods}. Dynamic analysis and DSE do not answer the question
because they only consider a {\it finite number of paths} while infeasibility is about 
considering {\it all paths}. Also, (standard) syntactic static analysis is too
easily fooled by unknown patterns. Finally,  while recent semantic static analysis
approaches \cite{BalakrishnanR10,BardinHV11,KinderV10,SeppMS11} can in principle
address infeasibility questions, they are currently neither scalable nor robust
enough.

At first sight infeasibility is a  simple mirror of feasibility, however from an
algorithmic point of view they are not the same problem. Indeed, since solving
feasibility questions on general programs is undecidable, practical approaches
have to be one-sided, favoring either feasibility (i.e.~answering {\it ``feasible''}
or {\it "I don't know''}) or infeasibility (i.e.~answering  {\it "I don't know''}
or {\it ``infeasible''}). While there currently exist robust methods for answering
feasibility questions on heavily obfuscated codes, no such method exist for
infeasibility questions.

%% In the
%% meantime, formal methods like abstraction interpretation or symbolic
%% execution, have provided very good results over the last few years for software
%% testing at binary-level.
%% Symbolic execution have successfully been used for path coverage on
%% genuine and well-structured programs for bug-finding, Control-Flow Intergrity
%% but less often on highly obfuscated programs the topic of this paper.

\myparagraph{Goal and challenges}  
In this article, we are interested in solving automatically infeasibility questions
occurring during the reversing of (heavily) obfuscated programs. The intended approach
must be {\it precise} (low rates of false positives and false negatives) and  
able to {\it scale} on realistic codes both in terms of size ({\it efficient}) and
protection -- including self-modification ({\it robustness}), and {\it generic} enough
for addressing a large panel of infeasibility issues. Achieving all these goals
at the same time is particularly challenging. 
%For example, semantic static analysis approaches \cite{BalakrishnanR10,BardinHV11,KinderV10,SeppMS11} 
% can in principle answer to infeasibility questions, yet 
% these approaches are currently  neither scalable nor robust. % enough. % (especially to self-modification). 
% 
%As another example, DSE is robust and scalable, yet as already noticed, it cannot answer to infeasibility queries. 

\myparagraph{Our proposal}
We present {\it Backward-Bounded Dynamic Symbolic Execution} (\bbdse), the first
precise, efficient, robust and generic method for solving infeasibility questions.
To obtain such a result, we have combined
in an original and fruitful way, several state-of-the-art key features of formal
software verification methods, such as deductive verification \cite{Leino05},
bounded model checking \cite{BiereCCZ99} or DSE. Especially, the technique is
{\it goal-oriented} for precision, {\it bounded} for efficiency and combines
{\it dynamic information and formal reasoning} for robustness.

%% Hence, instead of applying the DSE forward another approach is to
%% start from the goal location and applying the DSE backward. In the
%% worst case, it is the same complexity than forward but the number
%% of backward steps can be bounded~\cite{all_backward}. By bounding
%% the backward steps, the path is restricted to a sub trace which not
%% fit necessarily well at solving the same kind of problems. Not only
%% in obfuscation we might be led to test either the feasibility or the
%% infeasibility of an event. While doing DSE backward if a property
%% is proved to be infeasible then, the property is an invariant as
%% everything that happened beyond  the bound is over-approximated.
%% The challenge in this paper is to show the efficiency and the 
%% well-funded of this algorithm by deobfuscating some technics that
%% prevent the disassembly and the reverse-engineering of  a program.
%% The two obfuscations addressed opaque predicates and call stack
%% tampering that fits the infeasibility criteria of the algorithm.

%\todo[inline]{
% scaling issue\\
% forward () vs backward\\
% feasible $\neq$ infeasible
%}

\myparagraph{Contribution}
The contribution of this paper are the following:

\begin{itemize}
    \item First, we highlight the importance of infeasibility issues in reverse
       and the urging need for automating the investigation of such problems.
       Indeed, while many deobfuscation-related problems can be encoded as
       infeasibility questions (cf.~Section \ref{sec:encoding})
       it remains a blind spot of state-of-the-art disassembly techniques. 

	\item Second, we propose the new {\it Backward-Bounded DSE} algorithm for
	   solving infeasibility queries arising during deobfuscation (Section \ref{sec:bb-DSE}). 
       The approach is both precise (low rates of false positives and false negatives),
       efficient and robust (cf.~Table \ref{tab:dse_vs_bbdse}), and it can address
       in a generic way a large range of deobfuscation-related questions -- for 
       instance opaque predicates, call stack tampering or self-modification 
       (cf.~Section \ref{sec:encoding}).     
       The technique draws from several separated advances in software verification,
       and combines them in an original and fruitful way. We present the algorithm
       along with its implementation within the \binsec\ open-source platform
       \footnote{http://binsec.gforge.inria.fr/} \cite{BINSEC15,SANER}.  

	\item Third, we perform an extensive experimental evaluation of the approach, focusing  on two  standard obfuscation schemes, namely   
            {\it opaque predicates}  and {\it call stack tampering}. 
In a  set of {\it controlled experiments with ground truth} based on open-source obfuscators  
(cf.~Section \ref{sec:control-xp}), we demonstrate that our method is very precise and efficient. 
           % allows to detect  opaque predicates with a very low false positive rate ($<3\%$), to identify genuine {\tt \small call-ret} pairs and to finely characterize 
           % malicious pairs. 
%
%\todo[inline]{attention, bien mettre ça dans le reste du ppaier}We also demonstrate that the approach significantly overcome some existing work (based on standard DSE) in terms of efficiency or precision (false positive). 
% 
Then, in a {\it large scale experiment with standard packers} (including self-modification and other advanced protections),   the technique is shown to scale on realistic obfuscated codes, 
both in terms of efficiency and robustness (cf.~Section \ref{sec:control-xp}).  
%

% \todo[inline]{insister sur super new for call stack ??}

 \item Finally, we present two practical applications of Backward-Bounded DSE.  
First, we describe an {\it in-depth case-study of the government-grade malware} \xtunnel \cite{calvet_visiting_2016} (cf.~Section \ref{sec:xtunnel}), where \bbdse allows to identify and 
remove all obfuscations (opaque predicates).   
We have been  able to automatically extract a de-obfuscated version of functions -- discarding almost 50\% of dead and ``spurious'' instructions,  and providing an insights 
into its protection schemes, laying a very good basis for further in-depth investigations.    
%
%
%
 %% practical benefits of bb-DSE, considering the standard obfuscation schemes of 
 %%            {\it opaque predicates} (Section  \ref{sec:po}) and {\it call stack tampering} (Section \ref{sec:callret}). Especially, bb-DSE allows to detect  
 %%            opaque predicate with a very low false positive rate ($<3\%$), it is able to identify genuine {\tt \small call-ret} pairs and to finely characterize 
 %%            malicious pairs.  We also briefly discuss  other potential applications  to self-modification (Section \ref{sec:self-modif}). 
 %%               % new taxonomy ? 
%
 Second, %we give a glimpse of how  symbolic, dynamic and static  method can be combined together for better disassembly. 
%Especially, 
we propose {\it sparse disassembly} (cf.~Section \ref{sec:sparse}), a combination of Backward-Bounded DSE, dynamic analysis and   standard (recursive, syntactic) static disassembly   
allowing to  {\it enlarge}  dynamic disassembly {\it in a precise manner} --  
getting the best of dynamic  and static techniques, together with encouraging  preliminary experiments.       
%
% 
%Especially, sparse disassembly is the only disassembly approach being scalable, robust, safe and expected to be almost complete in practice (cf.~Table \ref{tab:cfg_robustness}).   

%A comparison of sparse disassembly with other families of disassembly techniques is given in Table \ref{}.    

%%  disassembly able to {\it complete}  dynamic disassembly {\it in a safe way}, hence 
%% getting the best of dynamic and static disassembly.   
%% % 

%% an insight, into a sparse disassembly algorithm taking in account results of the
%% 	two analyses to perform its disassembly in a more safe manner than classical recursive
%% 	or linear-sweep algorithms. \mynote{more ?}
\end{itemize}

%\todo[inline]{Merger contrib 4 et 5 ? }

\noindent{\it Our implementation and experimental data will be made available if the paper is accepted for publication. } 

%% \begin{itemize}
%% 	\item a novel scalable algorithm of backward bounded DSE formalized and discussed
%% 	along with the various specifities of its implementation within the \binsecse platform.
%% 	(Section \ref{sec:bb-DSE})

%% 	\item a scalable detection method for invariant opaque predicate with a very low false positive rate ($<3\%$)
%% 	and various metrics for evaluating the ``likeliness'' of a given predicate to be opaque.
%% 	(Section \ref{sec:po})

%% 	\item a new taxonomy for call stack tampering obfuscation characterizing the violation
%% 	being made or the quality a a genuine call-ret pair. A detection algorithm implementing
%% 	this algorithm is also discussed given that this obfuscation has never been addressed
%% 	using DSE algorithms. (Section \ref{sec:callret})

%% 	\item an insight, into a sparse disassembly algorithm taking in account results of the
%% 	two analyses to perform its disassembly in a more safe manner than classical recursive
%% 	or linear-sweep algorithms. \mynote{more ?}
%% \end{itemize}

%\todo[inline]{
%- scalable backward-bounded DSE (good at infeasibility proofs) +(optims?)\\
%- implem (in Binsec/SE)\\
%- sound \& complete\\
%- PO detection detection method (low FP, scalable..)\\
%- call/ret detection method (+ new taxonomy) (premiere fois DSE pour call/ret)\\
%- sparse disassembly (disass assisted with analysis results)
%}

\myparagraph{Discussion}
Several remarks must be made about the work presented in this paper. 

\begin{itemize}
\item First, while we
essentially consider opaque predicates and call stack tampering, \bbdse can also be 
useful in other obfuscation contexts, such as flattening or virtualization. Also
self-modification is inherently handled by the dynamic aspect of \bbdse.

\item Second, while we present one possible combination for sparse disassembly, other
combinations can be envisioned, for example by replacing the initial dynamic analysis by a (more complete)  
DSE \cite{ccs2015} or by considering more advanced static disassembly techniques \cite{binary_not_easy}. 

\item Finally, some recent works target opaque predicate detection  with  standard forward DSE~\cite{loop}. 
As already pointed out,  DSE is not  tailored to infeasibility queries, while \bbdse is --  % adapted to infeasibility queries, whi 
cf.~Sections \ref{sec:control-xp} and \ref{sec:related}.  % supporting this claim. . 

\end{itemize}

%some works are implicitely targeting infeasibility queries and especially opaque predicates~\cite{loop}.
%Unfortunately, by doing it in a forward manner they needlessly have to deal with the whole path predicate
%for each predicates to check. As consequence they make use of taint to counterbalance
%which far from being perfect brings additional problems (under-tainting/over-tainting).
%Thus, while backward-bounded DSE seems to be the most appropriate way to solve infeasibility problems
%no researches have used this technique.

\myparagraph{Impact} 
Backward-Bounded DSE does not supersede existing disassembly approaches, it complements
them by addressing infeasibility questions. Altogether, this work paves the way
for robust, precise and efficient disassembly tools for obfuscated binaries,
through the careful combination of static/dynamic and forward/backward approaches.

\begin{table}[htbp]
\caption{Disassembly methods for obfuscated codes}\label{tab:dse_vs_bbdse} \vspace{2mm}
%\resizebox{\columnwidth}{!}{
\def\arraystretch{1.2}

\begin{tabular}{|l|c|c|c|c|c|} 
\hline
	             & feasibility 				 & infeasibility & \multirow{2}{*}{efficiency} & \multirow{2}{*}{robustness} \\
	             & query       				 & query         &            &            \\
\hline
dynamic analysis & \good/\bad $^{(\dagger)}$ & \bad   		 & \good 	  & \good      \\
\hline
DSE              & \good 					 & \bad   		 & \bad 	  & \good      \\
\hline
static analysis  & \multirow{2}{*}{\good}	 & \multirow{2}{*}{\good/\bad $^{(\dagger\dagger)}$} & \multirow{2}{*}{\good} & \multirow{2}{*}{\bad} \\
(syntactic)      &                         	 &               &            &            \\
\hline
static analysis  & \multirow{2}{*}{\bad}	 & \multirow{2}{*}{\good} & \multirow{2}{*}{\bad}  & \multirow{2}{*}{\bad}  \\
 (semantic)      &                         	 &               &            &            \\
\hline
\hline
\bbdse           & \bad 					 & \ \ \good $^{(\ddagger)}$  & \good & \good  \\
\hline
\end{tabular}

$(\dagger)$: follow only a few traces %, hence may miss subtle behaviors

$(\dagger\dagger)$:  very limited reasoning abilities

$(\ddagger)$: can have false positive and false negative, yet very low in practice        % false negative: k not long enough.   false positive: pas assez de chemins dans le pre_k
\end{table}

\section{Background} \label{sec:background}

\myparagraph{Obfuscation} These transformations \cite{surreptitious} aim at hiding the real program behavior. While approaches such as virtualization or junk insertion 
make  instructions more complex to understand, other approaches directly hide the legitimate instructions of the programs -- making the reverser (or the disassembler) 
missing essential parts of the code while wasting its  
time in dead code. The latter category includes for example code overlapping, self-modification,  opaque predicates and call stack tampering. 

We are interested here in this latter category. For the sake of clarity, this paper 
mainly focuses on {\it opaque predicates} and {\it call stack tampering}.
% that lead disassembler to misbehave. 

\begin{description} 
\item[An opaque predicate] \qquad \qquad \qquad always
evaluates to the same value,  and  this property is ideally  difficult to 
deduce. The infeasible branch will typically lead the reverser (or disassembler) to a large and complex portion of useless junk code.     
Figure \ref{fig:po_example} shows the {\tt x86} encoding
of the opaque predicate $7y^2 - 1 \neq x^2$, as generated by O-LLVM~\cite{ollvm}. This
condition is always false for any values of \textsc{ds:x}, \textsc{ds:y}, 
so the conditional jump {\tt \small jz $<$addr\_trap$>$} is never going to be taken.

\item[A (call) stack tampering,] \qquad \qquad \qquad \qquad    or \call/\ret violation, consists in breaking the assumption
that a \ret instruction returns to the instruction following the call ({\it return site}), as exemplified in Figure \ref{fig:toy_tampering}.   
The benefit is twofold: the reverser might be lured into exploring useless code starting from the return site, 
while the real target of the \ret instruction will be hidden from static analysis. 
%
%% Figure \ref{fig:toy_tampering}
%% shows the trace of a stack tampering example found in the Aspack packer\footnote{www.aspack.com}.  
%% The second \call never return to its return site because it was poped in between.
%
%
%
%% In this situation, the disassembler often wrongly consider that the \ret
%% belongs to the second \call and try to disassemble its return site while it
%% is not reachable. 

\end{description}

\begin{figure}[htbp]
%\begin{scriptsize}
\begin{lstlisting}[escapechar=@]
mov   eax, ds:x
mov   ecx, ds:y
imul  ecx, ecx
imul  ecx, 7
sub   ecx, 1
imul  eax, eax
cmp   ecx, eax
jz    <addr_trap> //false jump to junk 
....  ........    //real code 
\end{lstlisting}
%\end{scriptsize}
\caption{opaque predicate: $7y^2 - 1 \neq x^2$}\label{fig:po_example}
\end{figure}

\begin{figure}[htbp]
%\resizebox{\columnwidth}{!}{
\centering
\def\arraystretch{1.2} {\footnotesize \tt 
\begin{tabular}{l|l} 
 <main>: & <fun>: \\
\hline
call <fun>                   & $[...]$ \\
.....  // return site   & push X \\
.....  // junk code     & ret //jump to X instead  \\
.....  // junk code     & \ \ \ \  //of return site \\
\end{tabular}
%}
}
\caption{Standard  stack tampering}\label{fig:toy_tampering}
\end{figure}

\myparagraph{Disassembly} 
We call {\it legit}, an instruction in a binary if it is executable in practice. 
Two qualities expected for a disassembly are (1) {\it soundness: does the algorithm recover only legit  instructions?} and (2) {\it completeness: does the algorithm recover all legit instructions?}  
Standard disassembly approaches essentially include (static) recursive disassembly, (static) linear sweep and  dynamic disassembly. 
\begin{description}

\item[Recursive disassembly] \qquad \qquad \qquad \ \ consists in exploring the executable file from a given (list of) entry point(s), recursively following the possible successors of each instructions. 
This technique may miss a lot of instructions, typically due to computed jums ({\tt \small jmp eax}) %-- more advanced approaches may rely on dedicated patterns \cite{xx}, 
or self-modification. In addition, the approach is easily fooled into disassembling junk code obfuscated by opaque predicates or call stack tampering. As such, the approach is neither safe nor complete. %%% even nice code
\item[Linear sweep] \qquad  \ \ \   consists in decoding linearly all possible instructions in the code sections. The technique aims at being more complete than recursive traversal, yet it comes at the price 
of many additional misinterpreted code instructions. Meanwhile, the technique can still miss instructions  hidden by code overlapping or self-modification. Hence the technique is unsafe, and incomplete on obfuscated codes.   

\item[Dynamic disassembly] \qquad \qquad \qquad \  retrieves only legit instructions and branches observed at runtime on one or several executions.  The technique is safe, but potentially highly incomplete -- yet, it does 
recover part of the instructions masked by self-modification, code overlapping, etc.  
 \end{description} 

For example, while \objdump is solely based on linear sweep, \IDA performs a
combination of linear sweep and recursive disassembly (geared with heuristics).

\myparagraph{Dynamic Symbolic Execution} 
Dynamic Symbolic Execution (DSE)~\cite{CS13,GLM12} (a.k.a {\it concolic execution})
is a formal technique for exploring program paths in a systematic way.  For each  path $\pi$, the
technique computes a symbolic \textit{path predicate} $\Phi_{\pi}$ as a set of constraints
on the program input leading to follow that path at runtime. Intuitively, $\Phi_{\pi}$ is the conjunction  
of all the branching conditions  encountered along $\pi$. 
This path predicate is then fed to an {\it automatic solver} (typically a SMT solver \cite{VanegueH12}). If a solution is found, it corresponds  
to an input data exercising the intended path at runtime.  
Path exploration is then achieved by iterating on all (user-bounded) program paths, and paths are discovered lazily thanks to an interleaving  of dynamic execution and symbolic reasoning \cite{DART,CUTE}. 
Finally, {\it concretization} \cite{DART,CUTE,issta2016} allows to perform relevant under-approximations of the path predicate by using the concrete information available at runtime. 

The main advantages of DSE  are  {\it correctness} (no false negative in theory, a bug reported is a bug found) and {\it robustness} (concretization does allow to handle unsupported features 
of the program under analysis without losing correctness). Moreover, the approach is easy to adapt to binary code, compared to other formal methods 
%. Especially,  many binary-level DSE tools have been developed 
\cite{BardinH11,GLM12,S2E,mayhem}.       
The very main drawback of DSE is the so-called {\it path explosion problem}: DSE is doomed to explore only a portion of all possible execution paths. As a direct consequence, DSE is incomplete  in the sense that it can 
only prove that a given path (or objective) is feasible (or coverable), but not that it is infeasible.

DSE is interesting for disassembly and deobfuscation since it enjoys the  advantages of dynamic analysis (especially, safe disassembly and robustness to self-modification or code overlapping), while being able to 
 explore a larger set of behaviors. Yet, while on small examples DSE can achieve  complete disassembly, it  often only slightly improves coverage (w.r.t.~pure dynamic analysis) 
on large and complex programs.

\section{Motivation} \label{sec:motivation}

Let us consider the obfuscated pseudo-code given in Figure~\ref{fig:motivating}.
The function {\tt \small <main>} contains an opaque predicate in \textcircled{1}
and a call stack tampering in \textcircled{2}.

\begin{figure}[htbp]
%\resizebox{\columnwidth}{!}{
\centering
\def\arraystretch{1.2} {\tt 
	\begin{tabular}{l|l} 
		 <main>:                        & <fun1>:            \\
		\hline
		if (C) \{ {\large \textcircled{1}}       & ..... 			 \\ 
		\ \ \ call <fun1> 				& push <X>{\large \textcircled{2}}\\
		\ \ \ //junk \textcircled{a}    & ret                \\
	    \}                              &                    \\
	    else \{                         & \\
	    \ \ \ call <fun2> \textcircled{b}  &                 \\
	    \}                              &                    \\
	    \cline{2-2}
	    //junk \textcircled{c}          & <fun2>:            \\
	    \cline{2-2}
	    ret //fake end of fun           &  .... \textcircled{d}\\
		<X>:                            &  ret               \\
		//payload                       &                    \\
	\end{tabular}
}
%}
\caption{Motivating example}\label{fig:motivating}
\end{figure}

Getting the information related to the opaque predicate and the call stack tampering
would allow to:
\begin{itemize}
	\item {\large \textcircled{1}} to know that {\tt <fun1>} is always called and reciprocally
	that {\tt <fun2>} is never called. As consequence \textcircled{b} and \textcircled{d}
	are dead instructions;
	
	\item {\large \textcircled{2}} to know that the {\tt \ret} of {\tt <fun1>} is
	tampered and never return to the caller. As consequence \textcircled{a} and 
	 \textcircled{c} are dead instructions. Such trick would also allow to hide
	 the real payload located at {\tt <X>}.
\end{itemize}

Hence the main motivation is not to be fooled by such infeasibility-based tricks
that slow-down the program reverse-engineering and its global understanding.

\begin{figure}[htbp]
	\resizebox{\columnwidth}{!}{
		\includegraphics{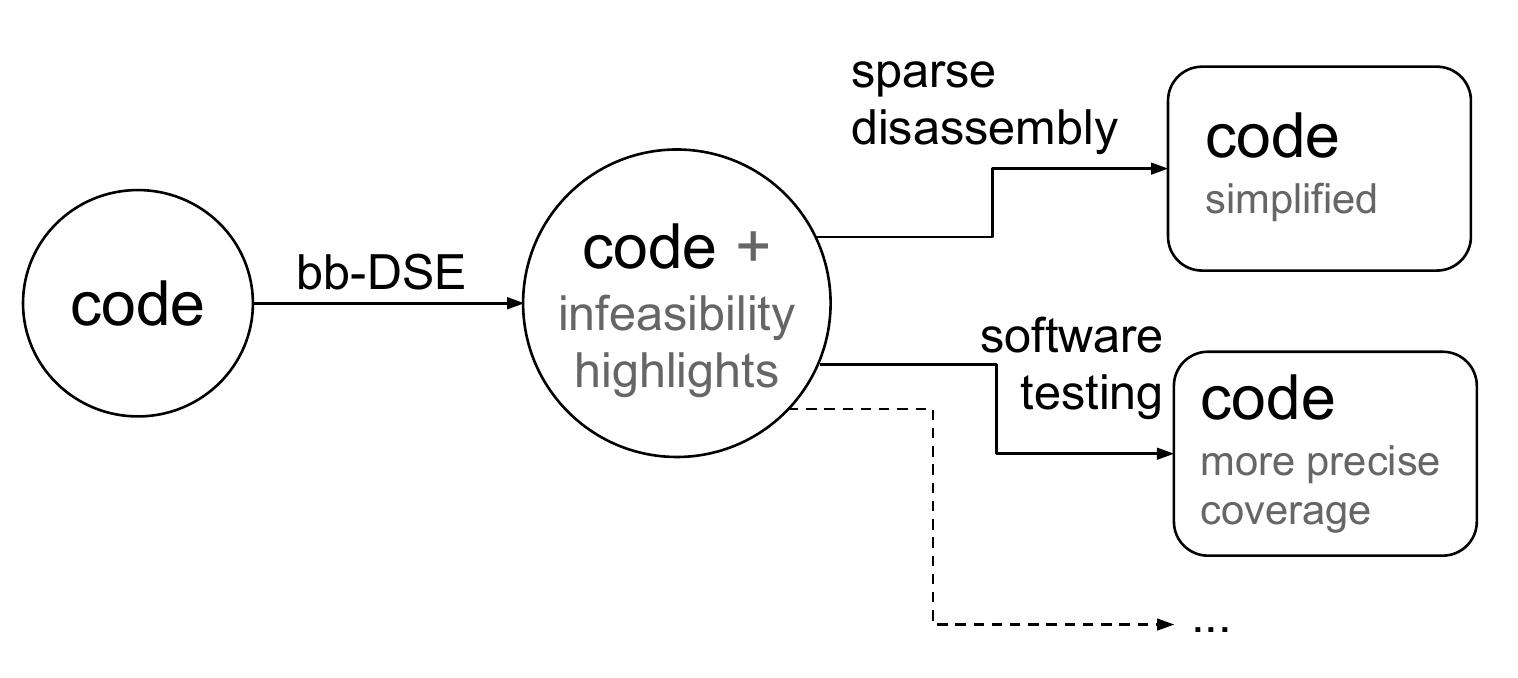}
	}
	\caption{motivation schema}\label{fig:motivation_schema}
\end{figure}

\myparagraph{Applications}
The main application is to improve a disassembly algorithm with such information, since 
 static disassembly will be fooled by such tricks and dynamic disassembly will
only cover a partial portion of the program. 
Our goal is to design an efficient method for solving infeasibility questions. This approach could then passes 
the original code annotated with infeasibility highlights  to other disassembly tools,  
which could take advantage of this information -- for example by avoiding disassembling dead instructions.   
Such a view is depicted in Figure \ref{fig:motivation_schema},  
%
%As none of the standard approaches
%is satisfactory proposing an automatic method based on bb-DSE could greatly leverage
%this information and to handle them as hints for disassembling the program.
and a  throughout study of such combination is discussed in ~Section \ref{sec:sparse}.

%\myparagraph{Motivation takeaway}
%Even though our main motivation is to improve the disassembly 
Moreover, such infeasibility
related information could also be used in other contexts, for instance  to obtain   more accurate
code coverage rates in software testing or to guide vulnerability analysis toward weak parts of the code. 

\section{Backward-Bounded DSE} \label{sec:bb-DSE}

%\todo[inline]{SB: make sure all points claimed in intro (tables) is explained here}

%\todo[inline]{key concepts, formalization, practical good properties, vs other techniques, choice of bound, implementation}

We present in this section the new Backward-Bounded DSE technique dedicated to solving 
infeasibility queries on binary codes.

\myparagraph{Preliminaries} We consider  a binary-level program $P$ with a given initial code address $a_0$. 
A state $s \triangleq (a,\sigma)$ of the program is defined by a code address $a$ and a memory state $\sigma$, 
which is a mapping from registers and memory to actual values (bitvectors, typically of size 8, 32 or 64).  
By convention, $s_0$ represents an initial state, i.e.~$s_0$ is of the form $(a_0,\sigma)$. 
The transition from one state
to another is performed by the $post$ function that execute the current instruction. 
%and update the location and path predicate accordingly. %\mynote{definir plus ?} 
%
%
An execution  $\pi$ is a sequence $\pi \triangleq (s_0 \cdot s_1 \cdot ... \cdot s_n)$, where $s_{j+1}$ is obtained by applying 
the $post$ function to $s_j$ ($s_{j+1}$ is the successor of $s_j$).

\smallskip 

Let us consider a predicate $\varphi$ over memory states.  We call {\it reachability condition} 
a pair   $c \triangleq (a,\varphi)$, with $a$ a code address. 
Such a condition $c$ is {\it feasible} if there exists a state $s \triangleq (a,\sigma)$ and an execution $\pi_s \triangleq (s_0 \cdot s_1 \cdot ... \cdot s)$ 
such that $\sigma$ satisfies $\varphi$, denoted  $\sigma \models \varphi$. 
It is said  {\it infeasible} otherwise. 
An {\it feasibility (resp.~infeasibility) question} consists precisely in trying to solve the feasibility (resp.~infeasibility) of such a reachability condition.

%

%By convenience, we will sometimes   call {\it predicate}  a reachability condition $(a,\varphi)$.  

%\todo[inline]{todo}   predecessor,  feasible, infeasible   

 %a path $\pi$ defined by $\pi \triangleq (s_0 \cdot s_1 \cdot ... \cdot s_n)$

%with $s_0 \in \mathbb{S}$ where $\mathbb{S}$ is the set of possibles states at each location defined by
%$\mathbb{S} \subseteq \mathbb{L}oc \times \Sigma$. 

%The transition from on state
%to another is performed by the $post$ function that execute the current instruction. 
%and update the location and path predicate accordingly. %\mynote{definir plus ?}

%% Similarly, the $pre$ function is defined by $pre\triangleq post^{-1}$. The backward
%% bounded slice of a path can be defined as $\pi_k \triangleq (s_i \cdot s_{i+1} \cdot ... \cdot s_n)$
%% where $n-k \leq i$. $post^{\leq k}$ and $pre^{\leq k}$ can be defined as the bounded application 
%% of the $post,pre$ function on the given path that respectively apply the transitions from $s_0 \mapsto s_k$
%% and $s_n \mapsto s_{n-k}$. 

%% Given a predicate $\varphi_p$ to check at location $l_p$.

%% Given the bound $k$, the predicate is not coverable if $\forall s \in \mathbb{S}, pre^{\leq k}(l_p, s) \in \emptyset$, meaning that 
%% there is no state $s \in \mathbb{S}$ where $s, l_p, \varphi_p \vdash SAT$.
%% As corollary, if $\varphi_p$ is not coverable for $k$, $\varphi_p$ is not coverable with $pre^{\leq k+1}$
%% as $pre$ is a monotonic function. The relative completeness is defined by
%% $\exists s\in \mathbb{S}, such\ as\ s,l_p, \varphi_p \vdash SAT \Rightarrow pre^{\leq k}(l_p, s) \notin \emptyset$.

\medskip 

These  definitions do not take self-modification into account. They can be extended to such a setting 
by considering code addresses plus waves or phases \cite{codisasm}.  

%self-modif, codisasm

\myparagraph{Principles} %\todo[inline]{Un peu de pédagogie sur le DSE...}
%% Forward DSE  performs well at  
%% finding new paths and generating new inputs, i.e.~answering . Its main
%% problem is scalability on large programs which is usually the
%% case with obfuscated code. 
We build on and combine 3 key ingredients from popular software verification methods: 
\begin{itemize}
\item backward reasoning from deductive verification, for {\it precise goal-oriented reasoning}; 
\item combination of dynamic analysis and formal methods (from DSE), for {\it robustness}; 
\item bounded reasoning from bounded model checking, for {\it scalability} and the ability {\it to perform infeasibility proofs}. 
\end{itemize}

The initial idea of \bbdse is to perform a  {\it \bf backward reasoning}, similar to the one of DSE but going from successors to predecessors (instead of the other way). %  of  
Formally, DSE is based on the $post$ operation while \bbdse is based on its inverse $pre$.  
%
%Perfect forward and backward reasoning $post^*$ and $pre^*$ (i.e.~fixpoint iterations of relations $post$ and $pre$) are symmetric, and both can be used to check feasibility and infeasibility 
%questions. 
%
Perfect backward reasoning $pre^*$ (i.e.~fixpoint iterations of relation  $pre$, collecting all predecessors of a given state or predicate) can be used to check feasibility and infeasibility 
questions. 
But this relation is  not computable.

Hence, we rely on computable {\it \bf bounded reasoning}, namely $pre^{k}$, i.e.~collecting all the ``predecessors in   $k$ steps'' ($k$-predecessors) of a given state (or predicate).  
%
% feinte : si arrive sur initial, reste dans k-pred quand même (init boucle sur lui-même)
%
% (for \bbdse)  $post^{\leq k}$ (for DSE)  and . 
Now symmetry does not hold anymore: 
% 
%  $post^{\leq k}$  (hence DSE) can answer positively to feasibility queries, but cannot falsify them,  
while $pre^{k}$  can answer {\it positively to infeasibility queries} (if a predicate has no $k$-predecessor, it has no  $k'$-predecessor for any $k'>k$  and cannot be reached), but {\it 
cannot falsify them}   
(because it could happen that a predicate is infeasible,  for a reason beyond the bound $k$).    
% 
%
%And indeed  $pre^{\leq k}$  works well at proving infeasibility since it over-approximates 
%the logical memory and register state of previous instructions (cf. Figure\ref{fig:prek}).    % in the trace.
%
Moreover, it is efficient as the computation does not depend on the program size or trace length, but on the user-chosen bound $k$.  
%
%
% The fully symbolic $pre^{\leq k}$ approach is {\it correct}: a query answered as infeasible is indeed infeasible;  but {\it incomplete}: it can now miss infeasible objectives  if $k$ is set too short.    
 %It can be applied beyond the obfuscation
%field and yield interesting results for vulnerability or safety purposes.
%

\medskip 

In practice, checking whether $pre^{ k} = \emptyset$ can be done in a symbolic way, like it is done in DSE: 
the set   $pre^{ k}$ is computed implicitly as a logical formula  (typically, a quantifier-free first-order formula over bitvectors and arrays), 
which is unsatisfiable iff the set if empty. This formula is then passed to an automatic solver, typically a SMT solver \cite{VanegueH12} such as \ZE.  

\medskip

Yet, backward reasoning is still very fragile at binary-level, since computing $pre$ in a perfect way  may be highly complex because of dynamic jumps or self-modification.  
 The last trick  is to combine this  $pre^{k}$ reasoning  with {\it \bf  dynamic traces}, so that the whole approach benefits from the robustness of dynamic analysis. 
Actually, the $pre^{k}$ is now computed w.r.t.~the control-flow graph {\it induced by a given trace} $\pi$ -- in a dynamic disassembly manner. 
We denote this {\it sliced} $pre^{k}$ by $pre_{\pi}^{k}$. 

\begin{figure}
\resizebox{\columnwidth}{!}{
\includegraphics{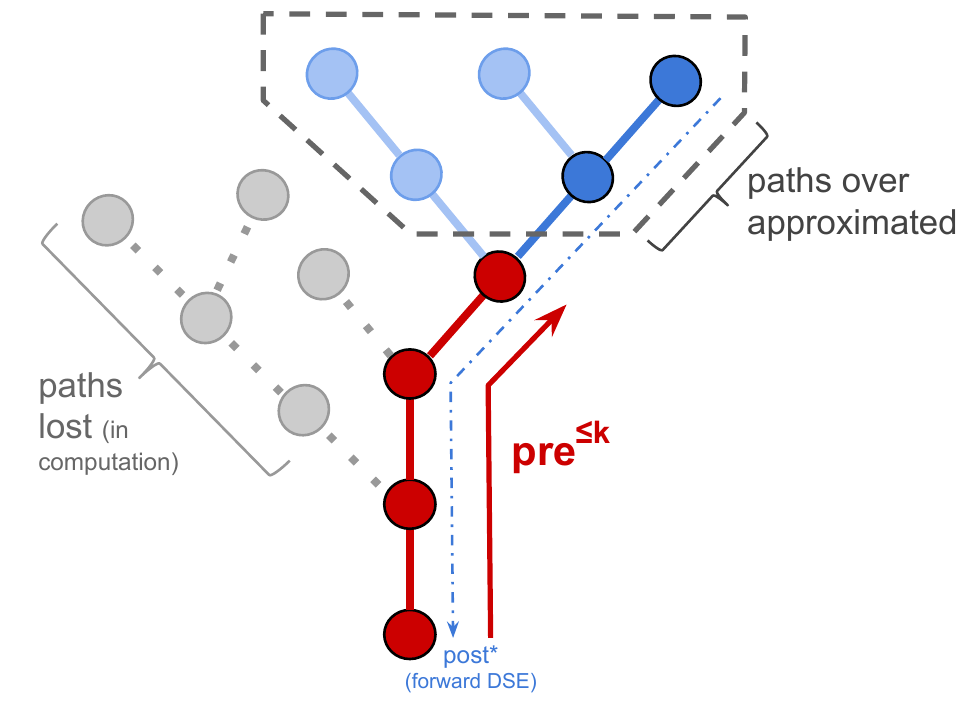}
}
\caption{$pre^k$ schema}\label{fig:prek}
\end{figure}

Hence we get {\it robustness}, yet since some real parts of $pre^{ k}$ may be missing from $pre_{\pi}^{ k}$, we {\it now lose correctness}: we may have false positive FP (because 
 $pre_{\pi}^{k}$ will be incomplete w.r.t $pre^{k}$), additionally 
 to the  false negative FN due to ``boundedness'' (because of too small $k$).  
A picture of the approach is given in Figure \ref{fig:prek}.

\myparagraph{Algorithm}  Considering a reachability condition $(a,\varphi)$,    \bbdse starts with a dynamic execution $\pi$: 
\begin{itemize}
\item if $\pi$ reaches code address $a$, then compute    $pre_{\pi}^{ k}((a,\varphi))$ as a formula and solve it
   \begin{itemize}
   \item if it is UNSAT, then the result is INFEASIBLE; 
   \item if it is SAT, then the result is UNKOWN; 
   \item if it is TO (timeout), then the result is TO; 
   \end{itemize} 
\item otherwise the result is UNKOWN.  
\end{itemize}

As a summary, this algorithm enjoys the following good properties: it is efficient (depends on $k$, not on the trace or program length) and as robust as dynamic analysis. 
On the other hand, the technique may report both false negative (bound $k$ too short) and  false positive (dynamic CFG recovery not complete enough). 
{\it Yet, in practice, our experiments demonstrate that the approach performs very well, with very low rates of FP and FN. } Experiments are presented in Sections 
\ref{sec:control-xp}, \ref{sec:packers} and \ref{sec:xtunnel}.

\smallskip

By convenience, we will not distinguished anymore between the predicate $\varphi$  and the reachability condition $(a,\varphi)$ if $a$ is clear from context.

 %In the following we do not distinguish anymore between $pre^{\leq k}$ and $pre_{\pi}^{\leq k}$, since it will be clear from context which one is, intended.   %\forjournal{bad idea !!} 

%% Recall that a (memory) state is a valuation of all variables and memory of the program, and an  invariant $Inv$ of a program $P$ 
%% is a property such that any reachable state $s$ satisfies $Inv$.  

%% Given an initial set of states $S_0$ and an invariant represented by a pair $(l,\psi)$, where $l$ is a code address and $\psi$ a formula, we have that 

%% \qquad $pre^*(\sem{(l,\psi)}) \cap S_0 = \emptyset$  % iff  $post^*(S_0) \cap \sem{(l,\psi)} = \emptyset$ 
%% iff
%%    $(l,\neg \psi)$ is an invariant of $P$ 

%% But computing  $pre^*$ is usually infeasible. Hence, we now {\it bound} the backward reasoning into $pre^{\leq k}$, meaning relationship $pre$ iterated $\leq k$ steps.  
%% % 
%% We have now the following relationship:  

%% \qquad $pre^k(\sem{(l,\psi)}) \cap S_0 = \emptyset \Rightarrow $ $(l,\neg \psi)$ is an invariant of $P$ 

%% The idea of bb-DSE
%% is to get rid of this issue by applying the DSE from 
%% the location of interest up in the trace until a given
%% bound. 

\myparagraph{Implementation}
This algorithm is implemented on top of \binsecse~\cite{SANER}, a forward DSE
engine inside the open-source platform \binsec~\cite{BINSEC15} geared to formal
analysis of binary codes. The platform currently proposes a front-end from
\texttt{\small x86} (32bits) to a generic intermediate representation called
DBA~\cite{CAV11} (including decoding, disassembling, simplifications). It also provides several
semantic analyses, including the \binsecse DSE engine \cite{SANER}. \binsecse
features a strongly optimized path predicate generation as well as highly configurable
search heuristics \cite{SANER,BardinHV11} and C/S policies \cite{issta2016}. The
whole platform amount for more than 40k of OCaml line of codes\footnote{{\small \tt http://binsec.gforge.inria.fr/tools}}.

\binsec also makes use of two other components. First, the dynamic instrumentation
called \pinsec, based on Pin %~\cite{} 
in charge to  run  the program and to record
all runtime values along with self-modification layers. Written in C++ it amounts
for more than 3k lines of code. Second, \idasec is an \IDA plugin written in Python
($\sim$13k loc) aiming at triggering analyzes and post-processing results generated
by \binsec.

The \bbdse algorithm  is tightly integrated in the \binsecse component. Indeed, when solving a  predicate feasibility, 
 \binsecse DSE performs a backward pruning pass aiming at removing any
useless variable or constraint. \bbdse  works analogously, 
but also takes into account the distance from the predicate to solve:  %between from the predicate to remove any definition.
any definition beyond  the $k$ bound is removed. In a second phase, the algorithm
creates a new input variable for any variable used but never defined in the
sliced formula. The $k$ bound value is defined by the user and can be
modulated as needed.

\section{Solving Infeasibility Questions with \bbdse} \label{sec:encoding}

We show in this section how several natural problems encountered during deobfuscation and disassembly 
can be thought of as infeasibility questions, and solved with \bbdse.  

\subsection{Opaque Predicates} \label{sec:po} \label{sec:po_approach} 

As already stated in Section \ref{sec:background},  an opaque predicate (OP)  is a predicate always evaluating to the same value. % to true or false
%which if inserted on a conditional jump would create a dead branch. 
They have successfully been used in various domains~\cite{Sok-diversity,surreptitious}. Recent
works~\cite{loop} identify three kinds of opaque predicates:
\begin{itemize}
	\item \textit{invariant}: always true/false due to the structure of the predicate itself, regardless of inputs values, 

	\item \textit{contextual}: opaque due to the predicate and its constraints on input values,  

	\item \textit{dynamic}: similar to contextual, but  opaqueness comes from dynamic properties on the execution (e.g.~memory).  
	%comparing two dynamic configuration that are made to be disjoint thanks to
	%dynamically computed values (eg: two disjoint pointer values etc..)
\end{itemize}

%% \begin{table}[htbp]
%% \caption{OP implemented in O-LLVM}\label{tab:po_ollvm}
%% \resizebox{\columnwidth}{!}{
%% \def\arraystretch{1.2}
%% \begin{tabular}{lll}
%% \hline
%%  \multicolumn{2}{l|}{Formulas} & Comment \\
%% \hline
%% $\forall x,y \in \mathcal{Z}$ & $y < 10 || 2|(x \times (x-1))$      & (the only in O-LLVM)\\
%% $\forall x,y \in \mathcal{Z}$ & $7y^{2} - 1 \neq x^{2}$             & \\
%% $\forall x \in \mathcal{Z}$   & $2|(x + x^{2})$                     & \\
%% $\forall x \in \mathcal{Z}$   & $2|\lfloor \frac{x^{2}}{2} \rfloor$ & ($2^{nd}$ bit of square always 0)\\
%% $\forall x \in \mathcal{Z}$   & $4|(x^2 + (x+1)^{2})$               & \\
%% $\forall x \in \mathcal{Z}$   & $2|(x \times (x+1))$                & \\
%% \hline
%% \end{tabular}
%% }
%% \end{table}

%% Table \ref{tab:po_ollvm} shows various invariant opaque predicates that we added
%% in O-LLVM to widen the predicates diversity inserted in programs.

%% %\todo[inline]{
%% %- Intro sur POs\\
%% %- arith etc, conditional/unconditional\\
%% %- exemple (faire tableau)
%% %}

\myparagraph{Approach with \bbdse} Intuitively, to detect an opaque predicate the idea
is to backtrack all its data dependencies and gather enough constraints to conclude to the infeasibility of the predicate.  
If the predicate
is local (invariant), the distance from the predicate to its input instantiation
will be short and the predicate will be relatively easy to break. 
 Otherwise (contextual, dynamic) 
  the distance is  linear with the trace length,  which does not necessarily
scale.

This is  a direct application of \bbdse, where $ p=(a,\varphi)$ is the pair address-predicate for which
we want to check for opacity. % -- recall that $a$ is the code address, and $\varphi$ the predicate itself.   
We call $\pi$ the execution trace under attention (extension to a set of traces is straightforward).  Basically, the detection algorithm is the following: 
\begin{itemize}
\item if     $p$ is dynamically covered by  $\pi$, then returns FEASIBLE;  
\item otherwise, returns \bbdse($p$), where INFEASIBLE is interpreted as ``opaque''. 
\end{itemize}

The result is guaranteed solely for FEASIBLE, since \bbdse has both false positives and false negatives. Yet, experiments (Sections \ref{sec:control-xp}, \ref{sec:packers}, \ref{sec:xtunnel}) show that 
these error ratios are very low in practice.

Concerning the choice of bound $k$, experiments in Section \ref{sec:control-xp} demonstrates that a value between 10 and 20 is a good choice with invariant opaque predicates. 
Interestingly, the \xtunnel case study (Section \ref{sec:xtunnel}) highlights that such rather small bound values may be sufficient to detect 
opaque predicates with long dependency chains (up to 230 in the study, including contextual opaque predicates), since we do not always need to recover all the information in order to conclude on the infeasibility. 

%  \ref{sec:xtunnel}

% detail : bound : 10-30 ok in practice (invariant), capture even longer ones (xtunnel, 230)

\subsection{Call Stack Tampering} \label{sec:callret}

Call stack tampering consists in   altering  the standard 
compilation scheme switching from function to function by associating a \call and a \ret and making
the \ret to return to the call next instruction. The \ret is tampered (a.k.a violated)
if it does not return to the expected return site pushed on the stack at the call.
%This paper propose a more precise characterization of a call stack tampering.

\myparagraph{New taxonomy}  
%From a semantic point of view a \ret is simply an indirect
%jump to the first value on the stack which lead to various
%cases of violation. 
In this work we refine the
definition of a stack tampering in order  to characterize it better. 
%For instance a it is
%important to distinguish if a \ret does not return well because the
%value has been overwritten or if the stack pointer has been
%tampered. This approach identifies the following \ret properties:
\begin{itemize}
	\item \textbf{\textit{integrity}}: does \ret return to the same address as pushed by the \call?
It characterizes if the tampering takes place or not.
% place or not using the dynamic values of the trace
%and validate them through DSE. 
A \ret  is then either \genuine (always returns to the caller) or \violated. 

%	\item \textit{strength}. can the \ret be tampered if it returned and respectively
%	can it return if it was tampered
	\item \textbf{\textit{alignment}}: is the stack pointer ({\tt esp}) identical at \call and \ret?
	If so, the stack pointer is denoted \alignedd, otherwise \disalignedd. 

	\item \textbf{\textit{multiplicity}}: in case of violation, is there only one possible \ret target? 
         This case is noted \single, otherwise \multiple. % Note that knowning if mulitple targets are possible
	%would require a forward DSE.
\end{itemize}

\myparagraph{Approach with \bbdse}  
%This study uses \bbdse to detect the
%tampering which of the knowledge of the authors have never been used. 
The
goal is to check several properties of the  tampering using \bbdse. We consider the following predicates on a \ret instruction: 
%
%in DSE by checking
%the validity of some predicates. A predicate $\varphi$ is VALID iff:
%$\varphi \vdash SAT \wedge \neg\varphi \vdash UNSAT$. To characterize the violation
%the following queries are performed:
%
\begin{itemize}

	\item $@[esp_{\{call\}}] = @[esp_{\{ret\}}]$: Compare the content of the 
	value pushed at call $@[esp_{\{call\}}]$ with the one used to return $@[esp_{\{ret\}}]$. 
	If it evaluates to VALID, the \ret  cannot be tampered \genuine. If it evaluates to UNSAT, a violation necessarily occurs \violated.  
        Otherwise, cannot characterize integrity. 
        % If SAT, the concrete values from
	% the trace should be used to characterize the \ret integrity (typically with forward DSE). 

	\item $esp_{\{call\}} = esp_{\{ret\}}$: Compare the logical \textsc{esp} value at the \call
	and at \ret. If it evaluates to VALID, the \ret necessarily returns at the same stack offset
	\alignedd, if it evaluates to UNSAT the \ret is \disalignedd. Otherwise cannot characterize
	alignment.

	\item $\mathcal{T} \neq @[esp_{\{ret\}}]$: Check if the logical \ret jump target $@[esp_{\{ret\}}]$ 
	can be different from the concrete value from the trace ($\mathcal{T}$). If it evaluates
	to UNSAT the \ret cannot jump elsewhere and is flagged \single.  Otherwise cannot characterize multiplicity.         %Note that \multiple cannot be found by our approach, unless several distinct (non-legit) targets are observed at runtime. 
\end{itemize}

The above cases can be checked by \bbdse (for checking VALID with some predicate $\psi$, we just need to query \bbdse with predicate $\neg \psi$).   
Then, our detection algorithm works as follow, taking advantage of \bbdse and dynamic analysis:  

\begin{itemize}
\item  the dynamic analysis can tag a \ret as:  \violated, \disalignedd, \multiple; 
\item  \bbdse can tag a \ret as: \genuine,  \alignedd, \single  (\violated and \disalignedd are already handled by dynamic analysis). 
\end{itemize}

As for opaque predicates, dynamic results can be trusted, while \bbdse results may be incorrect. 
Table \ref{tab:callret_taxonomy} summarizes all the possible situations.

\begin{table}[htbp]
	\caption{Call stack tampering detection}\label{tab:callret_taxonomy}
	\resizebox{\columnwidth}{!}{
		\centering \def\arraystretch{1.4}
		\begin{tabular}{|c|l|l|l|}
			\hline
			Runtime Status  & \multicolumn{1}{c|}{integrity}&  \multicolumn{1}{c|}{alignment}&  \multicolumn{1}{c|}{multiplicity} \\
			\hline
			 RT Genuine   &                         & RT: KO\disalignedd                    &                  \\
			              & VALID: \genuine         & - VALID: \alignedd                    &                  \\
%	    \mycomment{\genuine}        &                         & \mycomment{- UNSAT:  \disalignedd}    &                  \\
			\hline
			RT Tampered                &                         & RT:              KO\disalignedd       & RT: (2+)\multiple\\
			\violated     &                         & - VALID: \alignedd                    & - UNSAT: \single \\
			\hline
		\end{tabular}
	}
\end{table}

This call stack tampering analysis uses \bbdse,  
but with a slightly non-standard setting. Indeed, in this case the bound $k$  will be different
for every \call/\ret pair. The trace is analysed in a forward manner, keeping a formal  stack of \call instructions.  
Each \call encountered is pushed to the formal stack.
Upon \ret, the first \call on the formal stack is poped and \bbdse  
is performed,  where $k$ is the distance between
the \call and the \ret. 
% The different queries are then solved on this bounded path slice. 
%

From an implementation point of view, we must take care of possible  corruptions of the formal stack, which may happen for example in the following situations:   
%stack breaking the synchronicity of \call/\ret pairs:
\begin{itemize}
	\item Call to a non-traced function: because the function
	is not traced,   its \ret is not visible. In our implementation these calls are
	not pushed in the formal  stack; 
 
	\item Tail call~\cite{binary_not_easy} to non-traced function: tail calls consists in calling functions
	through  a jump instruction instead of \call to avoid stack tear-down. This is similar to the previous case, except that care must be taken in order to detect the tail call.  
	%Similarly, to the previous case a tail call in a non-traced
	%instruction will hide the \ret instruction, so the current call
	%stack head should be poped to keep synchronicity.
\end{itemize}

\subsection{Other deobfuscation-related infeasibility issues} 

\myparagraph{Opaque constant}
Similar to opaque predicates, opaque constants are expressions always evaluating
to a single value. % (while opaque predicate always evaluate to the same boolean).
%Moreover the goal is to make that property unclear for the reverser. 
Let us consider the expression $e$ and a value $v$ observed at runtime for $e$. Then, the opaqueness of $e$ 
reduces  to  the infeasibility of $e \neq v$. 
%If it is infeasible,  then the  expression is indeed constant. 

\myparagraph{Dynamic jump closure}
When dealing with dynamic jumps, switch, etc., we might be interested in knowing 
if all the targets  have been found. Let us consider a dynamic jump {\tt jump eax} 
for which 3 values $v_1, v_2, v_3$ have been observed so far. Checking the jump closure
can be done through checking the infeasibility of $eax \neq v_1 \wedge eax \neq v_2 \wedge eax \neq v_3$.

\myparagraph{Virtual Machine \& CFG flattening}
Both VM obfuscation and CFG flattening usually use a custom instruction pointer
aiming at preserving the flow of the program after obfuscation. In the case of
CFG flattening, after execution of a basic block the virtual instruction pointer
will be updated so that the dispatcher will know where to jump next. As such,
we can check that all observed values for the virtual instruction pointer have 
been found for each flattened basic block. Thus, if for each basic block we know
the possible value for the virtual instruction pointer and have proved it cannot
take other values, we can ultimately get rid of the dispatcher.

\myparagraph{A glimpse of conditional self-modification} \label{sec:self-modif}
Self-modification is a killer technique for blurring static analysis, since the
real code is only revealed at execution time. The method is commonly found in
malware and packers, either in simple forms (unpack the whole payload at once) or
more advanced ones (unpack on-demand, shifting-decode schemes~\cite{sok_packer}). 
The example in Figure \ref{fig:aspack_decoy} (page \pageref{fig:aspack_decoy})  taken from {\tt ASPack} combines an
opaque predicate together with a self-modification trick turning the predicate to
true in order to fool the reverser. Other examples from existing malwares have been
detailed in previous studies (NetSky.aa~\cite{ccs2015}).

Dynamic analysis allows to overcome the self-modification as the new modified code
will be executed as such. Yet, \bbdse can be used as well, to {\it prove interesting facts about self-modification schemes}. 
For example, given an instruction known to perform a self-modification,  we can take advantage
of \bbdse to know whether another kind of modification by the same instruction is possible or not (conditional self-modification).  
Let us consider an instruction 
{\tt $mov\ [addr],\ eax$} identified by dynamic analysis to generate some new code with value 
$eax=v$. Checking whether  the self
modification is conditional reduces to the   infeasibility of predicate $eax \neq v$.  
%An infeasible result would indicate the self-modification is not conditional.

%% A self-modification is {\it conditional} if the modification is triggered by some
%% user input (external API calls included). Dynamic execution can be tricked by
%% {\it conditional self-modification}, where a code-modifying instruction {\small
%% \tt store $@l_{addr},l_{val}$} can  write different instructions depending
%% on the logical value of the {\tt \small $l_{val}$}.

%% Given an instruction known to perform the self-modification we can take advantage
%% of \bbdse to know whether another modification as possible or not. Let's take
%% {\tt $mov [addr], eax$} identified by dynamic analysis to generate the new code and
%% having $eax=v$ we can encode the query $eax \neq v$ checking whether or not the self
%% modification is conditional or not. An \unsat result would indicates the self-modification
%% is not conditional.

As a matter of example, this technique has been used on the example of  Figure \ref{fig:aspack_decoy} 
to show that no other value than 1 can be written. This self-modification is thus
unconditional. 

%All this show, that \bbdse can be employed in a numerous ways in
%obfuscation in order to obtain some properties on the code under analyse.

%======================= OPAQUE PREDICATES =========================
%====================================================================

\section{Evaluation: Controlled Experiments} \label{sec:control-xp}

We present a set of controlled experiments with {\it ground truth values} aiming at evaluating 
the precision of \bbdse   as well as giving hints 
on its efficiency and  comparing it with DSE. %{\it All results of this section have been checked}. % or manually. 

\subsection{Preliminary: Comparison with Standard DSE} \label{sec:vs-dse}

We compare \bbdse with standard forward DSE, as well as with (unbounded) backward DSE. 
We are interested in comparing their efficiencies  and their adequacy to infeasibility questions -- through the distribution of their results,  between SAT, UNSAT and timeout.   
The experiment is performed on a trace  
of 115000 instructions and we check at each conditional jump if the branch
not taken is infeasible (UNSAT) or not (SAT), which is equal to checking if the branch is dead.
For \bbdse, we take the algorithm for opaque predicate detection described in Section \ref{sec:encoding},  
with   bound values  $k=100$ and $k=20$. We argue in latter experiments (Section \ref{sec:po}) that $k=20$ is a reasonable bound. 
We use the forward DSE of \binsecse, and backward DSE is obtained from \bbdse with a bound set to $\infty$. 
%
%We performed benchmarks to assess the scalability and robustness of the \bbdse
%algorithm applied on branch coverage. 
%
%% The benchmark is performed on a large path
%% of 115000 instructions. The benchmark checks for each conditional jump if the branch
%% not taken is infeasible, which is equal to checking if the branch is dead. The benchmark
%% is performed with forward DSE, backward DSE (unbounded) and \bbdse.
%
   
Results are presented in Table \ref{tab:bench_dse_vs_bbdse}. 
% shows for each, the total execution time, the
%number of predicates checked with theirs status and the number of timeouts (To=60s). 
%
While forward and backward  DSE provide similar results,  \bbdse clearly 
surpasses them in terms of efficiency,  spending less than a second for every predicate 
without any timeout ($\geq$ 2000 with DSE).  
%In comparison to unbounded backward DSE, the \bbdse provides a x58 speed-up. 
%
%  
From a result point of view, \bbdse with k=16 returns very few UNSAT answers compared to the other methods (54 vs $\geq$ 7000). Actually,  
this was expected since DSE takes the whole path into account, and while dead branches are rare in normal code, dead paths are very common.

%identifies  approximately the same number of \unsat branches and significantly more \sat branches
%which are likely timeout for DSE and backward DSE. 

\begin{table}[htbp]
	\caption{Benchmark DSE versus \bbdse}\label{tab:bench_dse_vs_bbdse}
	\resizebox{\columnwidth}{!}{
	\centering\def\arraystretch{1.2}
	\begin{tabular}{|l|c|c|c|c|c|} 
		\hline
					  & bound    & \multicolumn{3}{c|}{Cond. branch}       & Total \\
		\cline{3-5}
					  & k        & \#SAT      & \#UNSAT  & \#Timeout            & time  \\
		\hline
		forward DSE & /        & \mycomment{8322}  575 & {\bf 7749}    & {\bf 2460}               & \mycomment{63,780.27} {\bf 17h43m} \\
		\hline
		backward DSE  & $\infty$ & \mycomment{8321} 575 & 7748     & 2461               & \mycomment{64,080,84}  17h48m \\
		\hline
		\bbdse        & 100      & \mycomment{10784} 3378 & 7406   & 0                  & \mycomment{1,090.78}   18m78s \\
		\hline
		\bbdse        & 20       & \mycomment{10784} 10730 & {\bf 54}   & {\bf 0}                  & \mycomment{254.72}     {\bf 4m14s}  \\
		\hline
	\end{tabular}
	} 
\end{table}

%As conclusion results
%obtained validates the approach and emphasizes the scalability of \bbdse.

\myparagraph{Conclusion}  This preliminary experiment gives a clear demonstration on the advantages of \bbdse over DSE on infeasibility questions. 
Indeed, besides the dramatic gap in efficiency (which was of course expected since DSE depends on the whole size trace), DSE reports far more infeasible 
branches -- which would lead in practice to too many false positives. 
These results were  expected, as they are direct consequences of the design choices behind DSE  and \bbdse. On the opposite, \bbdse is not suitable for feasibility questions.

%   much more scalable : ok, was clearly expected
%    large diff in unsat answer : meaning that if DSE were used for , would conclude to much more ...   many false positive, because depend too much on the path, not on the predicate

\subsection{Opaque Predicates evaluation} \label{sec:po}

%\myparagraph{Research question} We are interested in the following questions: 
%\rquestion{1} what is the  detection capacity of our approach, and are the ratio of false positive and false negative acceptable?    
%
%\rquestion{2} is our approach efficient and scalable? 

We consider here the \bbdse-based algorithm for opaque predicate detection. We want to evaluate its precision, as well as to get insights 
on the choice of the bound $k$.

\myparagraph{Protocol and benchmark} We consider two sets of programs:  
%\begin{itemize}
%	\item 
(1) all 100 \coreutils without any obfuscation,  as a genuine reference data set, 
%	\item 
and (2) 5 simple programs taken from the State-of-the-Art
	in DSE deobfuscation~\cite{ccs2015} and obfuscated with  O-LLVM~\cite{ollvm}.  
%\end{itemize}
  Each of the 5 simple programs  was obfuscated
	20 times (with different random seeds) in order to balance the numbers of obfuscated samples and  genuine \coreutils. 
%(the non-deterministic
%	aspect of O-LLVM obfuscation makes all sample diversified completely different.
%
We have  added some new opaque predicates in O-LLVM (which is open-source) in order to maximize diversity (Table \ref{tab:po_ollvm}).

\begin{table}[htbp]
	\caption{OP implemented in O-LLVM}\label{tab:po_ollvm}
	\resizebox{\columnwidth}{!}{
		\def\arraystretch{1.2}
		\begin{tabular}{lll}
			\hline
			\multicolumn{2}{l|}{Formulas} & Comment \\
			\hline
			$\forall x,y \in \mathcal{Z}$ & $y < 10 || 2|(x \times (x-1))$      & (initially present in O-LLVM)\\
			$\forall x,y \in \mathcal{Z}$ & $7y^{2} - 1 \neq x^{2}$             & \\
			$\forall x \in \mathcal{Z}$   & $2|(x + x^{2})$                     & \\
			$\forall x \in \mathcal{Z}$   & $2|\lfloor \frac{x^{2}}{2} \rfloor$ & ($2^{nd}$ bit of square always 0)\\
			$\forall x \in \mathcal{Z}$   & $4|(x^2 + (x+1)^{2})$               & \\
			$\forall x \in \mathcal{Z}$   & $2|(x \times (x+1))$                & \\
			\hline
		\end{tabular}
	}
\end{table}

%
%
 %% In order to perform benchmark, we have choosen
%% the obfuscator O-LLVM~\cite{ollvm} in which we implemented various opaque predicates
%% (cf.~Table \ref{tab:po_ollvm}). O-LLVM is used because its core engine is
%% open-source, extensible and it already implement opaque predicate insertion mechanism.
%% Other obfuscation engines were not open-source~\cite{Collberg_tigress} or outdated.
%% In order to evaluate the Backward-Bounded DSE, two sets of programs were used:
%
%
 In total, 200 binary programs were used. For each of them a dynamic
execution trace was generated with a maximum length of 20.000
instructions. By tracking where opaque predicates were added in
the obfuscated files,  we are able a priori to know if a given 
predicate is opaque or not, ensuring a {\it ground truth evaluation}. Note  that we consider all predicates in \coreutils
to be genuine. % (and then opaque are then considered as  false positive). 
%This large but controlled experiment allowed to tests the accuracy of results. 
The 200 samples sums up a total of 1,091,986
instructions trace length and 11,725 conditional jumps with 6,170
genuine and 5,556 opaque predicates. 
%
%Usage of O-LLVM~\cite{ollvm}
%\todo[inline]{
%- Mix genuine programs, obfuscated programs (p.6-7) (envirronement controlé / same opaque/genuine ..)\\
%- opaque taken from previous DSE SOTA (debray)\\
%- valide car les traces des po était bien plus longue que celles de coreutils.\\
%- OLLVM ce qu'il implemente ce que j'ai ajouté\\
%- setup
%}
%
%\todo[inline]{
%explain my heuristic to know if a jmp is opaque or not ?   SB > no
%}
%
Finally, experiments were carried using different values for the bound $k$, and with a 5 second timeout per query. 

%% in
%% order to find which one fits best at finding all opaque predicates 
%% without yielding too much false positives. As a matter of estimation
%% a predicate marked opaque with $k=2$ is more likely to 
%% be opaque than a predicate marked opaque with $k=30$ since the path
%% constraint might make the predicate evaluate to UNSAT.

\begin{figure}
\caption{Graph opaque predicate detection ratio}\label{fig:po_graph}
\resizebox{\columnwidth}{!}{
\includegraphics{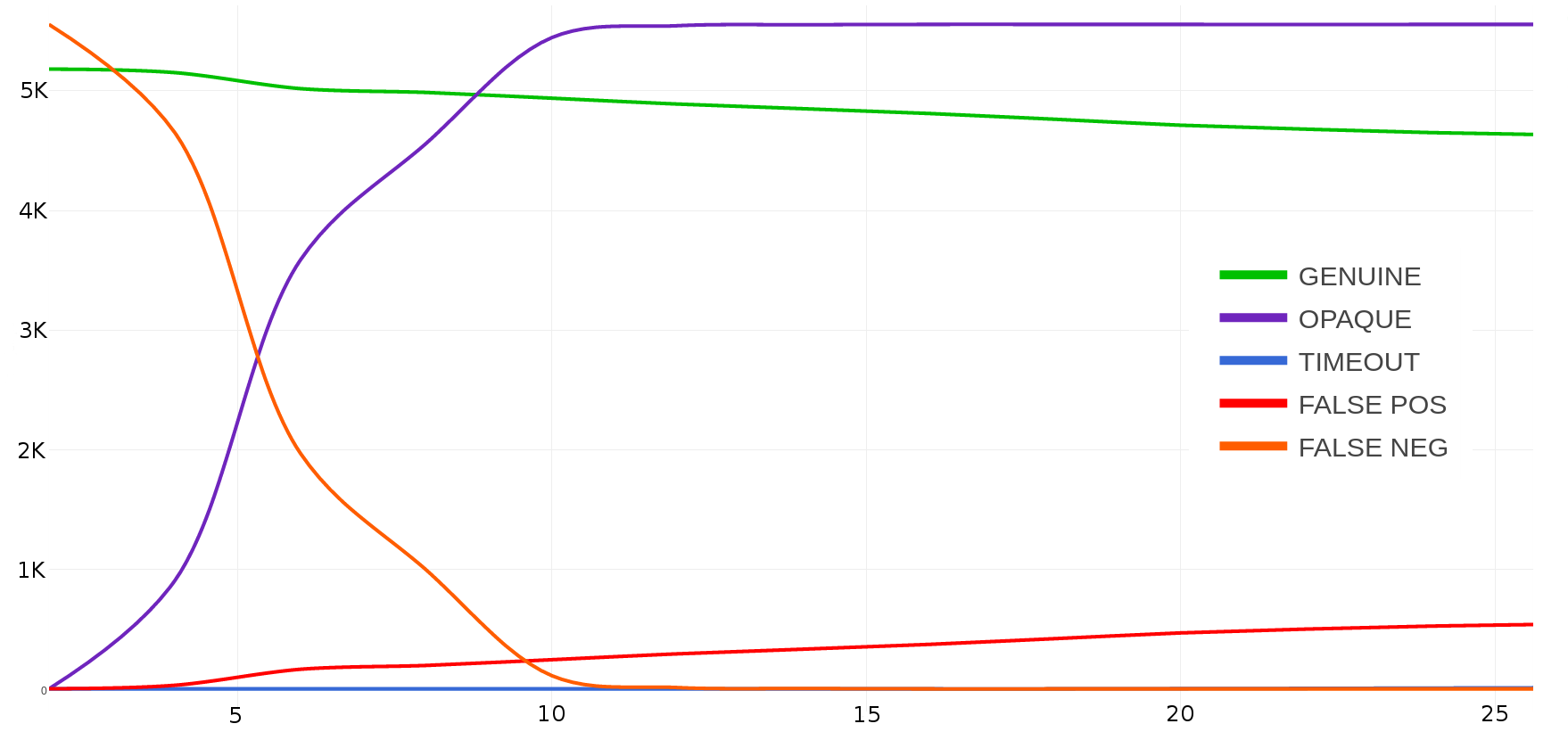}
}
\end{figure}

\myparagraph{Results} 
 Among the 11,725 predicates, 987 were fully covered
by the trace and were excluded from these results, keeping 10,739 predicates (and 5,183 genuine predicates). 
Figure \ref{fig:po_graph} and Table \ref{tab:po_results} show 
the relation between the
number of predicates detected as opaque (OP) or genuine (OK) as well as false positive (FP) and false negatives (FN) depending of the
bound value $k$. 
%The two last columns show the percentage of false positive
%among the number of all predicate and the percentage among opaque predicates.
The experiment 
shows a tremendous peak of opaque detection with 
$k=10$. Alongside, the number of false negative steadily decreases as the number of
false positive grows. 
An optimum is reached for $k=16$, with  no false negative, no timeout and a small number of  false positive (293), representing 6.28\% of all predicates
marked opaque and only 3.17\% of all predicates. In that case, the detection method achieves 1.46 false positive per sample (very low). 
Results are still very precise up to $k$ = 30, and very acceptable for $k$ = 50.

\begin{table}[htbp]
\caption{Opaque predicate detection results}\label{tab:po_results}
\resizebox{\columnwidth}{!}{
\def\arraystretch{1.2}
\begin{tabular}{|l|c|c|c|c|c|c|c|}
\hline
     \multirow{2}{*}{k} & \multicolumn{2}{c|}{OP (5556)} & \multicolumn{2}{c|}{Genuine (5183)} &  &  &  \\
\cline{2-5}
	   & ok    & miss & ok   & miss & TO & FP/Tot   & FP/OP \\
	   &       & (FN) & ok   & (FP) &  & (\%)   & (\%)     \\
\hline
  2        &  0    & 5556 & 5182 & 1    & 0  &  0.01   & 0.02       \\
  4        &  903  & 4653 & 5153 & 30   & 0  & 0.26    & 3.22       \\ 
  8        &  4561 &  995 & 4987 & 196  & 0  & 1.67    & 4.12        \\ 
  12        &  5545 & 11   & 4890 & 293 & 0  & 2.50    & 5.02       \\ 
  \bf 16  &  \bf  5556 &  \bf  0   &  \bf 4811 &  \bf  372  &  \bf 0  &  \bf 3.17   &  \bf  6.28      \\ 
%  \bf 16   &       &      &    &    &   &     &        \\ 
  \bf   20       &    \bf 5556 &   \bf 0    &   \bf 4715 &    \bf 468  &   \bf 2   &   \bf  3.99    &   \bf   7.77       \\ 
  \bf  24        &   \bf  5556 &   \bf 0    &   \bf 4658 &    \bf 525  &   \bf 7   &    \bf 4.48    &    \bf  8.63    \\ 
 32        &  5552 & 4    & 4579 &  604  & 25  &  5.15    &   9.81  \\ 
 40        &  5548 & 8    & 4523 &  660  & 39  &  5.63    &   10.63 \\ 
 50        &  5544 & 12   & 4458 &  725  & 79  &  6.18    &  11.56 \\ 
\hline
\end{tabular}
}

\smallskip 

. Timeout: 5 sec  

. a TO counts as a UNKNOWN result (hence, classify the predicate as genuine)

. 10,739 predicates, 5,556 opaque predicates, 5,183 genuine predicates

\end{table}

\myparagraph{A glimpse at efficiency} Taking  the same detection algorithm and  queries, we want to evaluate   
predicate solving time. 
% to check if
%it scales on a large number of predicates (here more than 10000).
Each predicate formula is solved using Z3\footnote{http://github.com/Z3Prover/z3} with a 5 seconds
timeout. Table \ref{tab:solving_time} gives for each value of $k$ 
  the total time taken for solving and the average time  per query.
For $k$ = 16 the average time per query is 0.018s which tends to prove
that this technique scales. Previous works aiming at solving invariant
opaque predicates~\cite{loop} (based on forward DSE) reports an  average of 0.49s per queries (min:0.09, max:0.79). 
Hence, \bbdse seems to  provides a very significant 
speed up. % (more than x27).  %\todo[inline]{Refais peut-etre les calculs pour etre sur..}

\begin{table}[htbp]
\caption{Solving time (10,739 queries)}\label{tab:solving_time}
%\resizebox{\columnwidth}{!}{
\def\arraystretch{1.2}  \scriptsize
{\hfill \begin{tabular}{|l|r|r|}
\hline
     k &  Total time (s)   & Avg/query(s) \\
\hline
2   &  89  &  0.008 \\
4   &  96  &  0.009 \\
%6  &  110  &  0.010 \\
8   &  120  &  0.011 \\
%10 &  136  &  0.013 \\
12  &  152  &  0.014 \\
%14 &  171  &  0.016 \\
16  &  197  &  0.018 \\
%18 &  221  &  0.021 \\
20  &  272  &  0.025 \\
%22 &  326  &  0.030 \\
24  &  384  &  0.036 \\
%26 &  444  &  0.041 \\
%28  &  533  &  0.050 \\
%30 & 617  &  0.058 \\
32  &  699  &  0.065 \\
%34 &  790  &  0.074 \\
%36  &  875  &  0.082 \\
%38 &  1018  &  0.095 \\
40  &  1145  &  0.107 \\
%42 &  1306  &  0.122 \\
%44  &  1465  &  0.136 \\
%46 &  1636  &  0.152 \\
%48  &  1889  &  0.176 \\
50 &  2025  &  0.189 \\
\hline
\end{tabular}
\hfill }
%}
\end{table}

%\todo[inline]{Faire benchmark temps résolution ? (pour humilier loop ?)}

\myparagraph{Conclusion}  \bbdse is very precise at detecting opaque predicates, and while a bound is of importance, the technique 
still performs very well if the bound is not too far from the optimal one ($k$ between 10 and 30). Query solving time is also very low.

\subsection{Call Stack Tampering evaluation} \label{sec:callret}

We want to assess the precision of the \bbdse  algorithm for stack tampering detection (Section \ref{sec:callret}). 
We consider two sets of programs: 
%\begin{itemize}
%\item  
(1) the 5 samples used  Section \ref{sec:po}, obfuscated with the source-to-source obfuscator Tigress~\cite{Collberg_tigress}; % (AntiBranchAnalysis transformation);   
%\item  
(2) 5 (randomly selected) \coreutils without any obfuscation. 
%\end{itemize}

%perform a controlled benchmark on obfuscated samples then similarly to OPs we perform a second
%benchmark on a few \coreutils to ensure it scales on real-world programs and
%do not yield false positives.

%We use the same dataset (5 samples) than in opaque predicates
%but this time, obfuscated with the source-to-source obfuscator Tigress~\cite{Collberg_tigress}.
Files have been obfuscated with the AntiBranchAnalysis transformation that replace
all conditional branches with calls and rets. The two schemes used by the obfuscator are:
\begin{itemize}
		\item push; call; ret; ret (the last \ret jumps on pushed value)
		\item push; ret (idem)
\end{itemize}

\myparagraph{Results} 
Results are given in Table \ref{tab:callret_results1}. The  5 obfuscated  
samples sums up 218 different \ret. Among them 77 are genuine and 141 violated (including 132 single-target).   
%Note that due to the homogeneity of schemes inserted by Tigress, the diversity of tags is
%very low.    
Our detection method achieves here a {\it perfect score}, with no false positive nor
false negative.  
%
%** *also : single targets, genuine, etc.  *****
%
On The 5 \coreutils, our method   does not yield {\it any false positive} and most of the 
\ret are  proved genuine (149/156). The few remaining unproved \ret come from 
unhandled libc side-effects, making formulas wrongly UNSAT.

\begin{table}[htbp]
\caption{Stack tampering results}\label{tab:callret_results1}
\centering
\resizebox{\columnwidth}{!}{
\def\arraystretch{1.2}
\begin{tabular}{|l||c|c|c||c|c|c|}
\hline
     \multirow{3 }{*}{Sample} & \multicolumn{3}{c||}{runtime genuine} & \multicolumn{3}{c|}{runtime violation} \\
\cline{2-7}
	        & \multirow{2}{*}{\#\ret $^{\dagger}$} & proved         & align/             & \multirow{2}{*}{\#\ret $^{\dagger}$} & alig/              & proved \\
                & \mycomment{uniq}         & genuine        & disal              & \mycomment{uniq}          & disal              & single       \\  
\hline
simple-if       & 6    &  6       & 6/0    & 9  & 0/0 & 8  \\
bin-search      & 15   &  15      & 15/0   & 25 & 0/0 & 24 \\
bubble-sort     & 6    &  6       & 6/0    & 15 & 0/1 & 13 \\
mat-mult        & 31   &  31      & 31/0   & 69 & 0/0 & 68 \\
huffman         & 19   &  19      & 19/0   & 23 & 0/3 & 19 \\
\hline
\hline
ls              & 30   & 30       & 30/0   & 0  & 0/0 & 0  \\
dir             & 35   & 35       & 35/0   & 0  & 0/0 & 0  \\
mktemp          & 21   & 20       & 20/0   & 0  & 0/0 & 0  \\
od              & 21   & 21       & 21/0   & 0  & 0/0 & 0  \\
vdir            & 49   & 43       & 43/0   & 0  & 0/0 & 0  \\
\hline

%ASPack          & 11   & 9       & 9/0    & 6  & 5/1  & 2  \\                           
%\hline
%ACProtect       & 0    & 0       & 0/0    & 48 & 45/1 & 45 \\ 
%\hline
%Crypter         & 125  & 94      & 94/0   & 78 & 0/30 & 32 \\ 
%\hline
%PE Lock         & 4    & 3       & 3/0    & 3  & 0/1  & 0  \\ 
%\hline
\end{tabular}
}

$^{\dagger}$each \ret\ is counted only once

\end{table}

\myparagraph{Conclusion} \bbdse performs very well here, with no false positive and a perfect score on obfuscated samples. 
The technique recovers both genuine \ret and single-source tampered  \ret.  
Interestingly, no tampered \ret were  
found on the few (randomly selected) \coreutils,  supporting the
idea that such tampering is not meant to occur in legitimate programs.

\subsection{Conclusion}

These different controlled experiments demonstrate  clearly that \bbdse is a very precise approach for solving different kinds of infeasibility questions. 
They also demonstrate that finding a suitable bound $k$ is not a problem in practice. Finally, the approach seems to be scalable. This last point will be definitely proved in Sections \ref{sec:packers} 
and  \ref{sec:xtunnel}.

%\todo[inline]{** conclusion sur small xp : much more efficient than DSE, precise}

%\todo[inline]{Pour ça il faudrait avec des résultats de résolution en avant ...}

%====================== Large scale XP  =====================
%===============================================================

\section{Large-scale Evaluation on Packers} \label{sec:packers}

To validate the scalability of \bbdse on representative codes, in terms of both size and protection, 
we perform a large scale experiment on packers with  the two detection algorithms already used 
in Section \ref{sec:control-xp}.  

\myparagraph{Context} Packers are programs embedding other programs and decompressing/deciphering them  at runtime. Since packers are   
used for software protection,   most of them contain several obfuscation schemes (including self-modification).   
As a matter of fact, packers are also widely used by malware, and actually in many cases they are  the only line of defense. % of many malware. 
Hence, {\it packers are very representative} for our study, both in terms of  malware protections  and size, as packed programs tend 
to have huge execution traces.

% As it is meant to
%provide protection, packers are more likely to contain obfuscation. 

\myparagraph{Protocol} We want to check if \bbdse\ is able to detect opaque predicates or call stack tampering on packed programs. 
For that, a large and representative set  of packers was chosen,  ranging from free to commercial
tools. 
Then a stub binary ({\tt hostname}) was packed by each packer.  
Analyses are then triggered on these packed programs  in a black-box manner, that is to say, without any 
prior knowledge of the internal working of the packers -- we do not know which obfuscation are used. 
For homogeneity, trace length are limited
to 10M instructions and packers reaching this limit were not analysed.

\subsection{Results}

Table \ref{tab:packers_partial} shows the partial results on 10 packers. The complete
results are given in Table \ref{tab:packers} in appendix. 
%
%\begin{itemize}
%\item 
First, \bbdse\ is efficient and robust enough to pass on most of the packed programs, involving traces of several millions of instructions and advanced protections such as 
self-modification.   
%
%\item 
Second, over the 32 packers,
420 opaque predicates and 149 call/stack tampering have been found, and many functions have also been proved genuine.   
All the results that have been manually checked appeared to be true positive (we did not checked them all because of time constraints).   
%Given the detection rate
%of opaque predicates and call stack tampering, checking all of them was too time-consumming, 
%yet all results manually checked appeared to be true positive. 
%\end{itemize}

\begin{table*}[htbp] 
\caption{Packer experiment OP \& Stack tampering}\label{tab:packers_partial}
\centering
\def\arraystretch{1.2}\small
%\resizebox{\textwidth}{!}{
\begin{tabular}{|l|r|r|c|c|c|c|c|}
\hline
%\multirow{3}{*}{Packers} &  & \multicolumn{1}{c|}{ } & \multicolumn{5}{c|}{Semantic infos} \\
%\cline{4-8}
\multirow{2}{*}{Packers}& \multirow{2}{*}{ size\ \ }& \multirow{2}{*}{tr.~len} & \multicolumn{3}{c|}{Opaque Pred.} & \multicolumn{2}{c|}{Stack tampering}\\
\cline{4-8}
                &           &         & Unk & OP & TO                     & OK (a/d/g)    & KO (a/d/s)      \\
\hline
ACProtect {v2.0}& 101K      & 1.8M    & 74  & 159& 0                      & 0(0/0/0)      & 48(45/1/45)         \\
ASPack {v2.12}  & 10K       & 377K    & 32  & 24 & 0                      & 11(7/0/7)     & 6(1/4/1)            \\
Crypter {v1.12} & 45K       & 1.1M    & 263 & 24 & 0                      & 125(94/0/94)  & 78(0/30/32)         \\
Expressor       & 13K       & 635K    & 42  & 8  & 0                      & 14(10/0/10)   & 0(0/0/0)            \\
nPack {v1.1.300}& 11K       & 138K    & 41  & 2  & 0                      & 21(14/0/14)   & 1(0/0/0)            \\
PE Lock         & 21K       & 2.3M    & 53  & 90 & 0                      & 4(3/0/3)      & 3(0/1/0)            \\
RLPack          & 6K        & 941K    & 21  & 2  & 0                      & 14(8/0/8)     & 0(0/0/0)            \\
TELock {v0.51}  & 12K       & 406K    & 0   & 2  & 0                      & 3(3/0/3)      & 1(0/1/0)            \\
Upack {v0.39}   & 4K        & 711K    & 11  & 1  & 0                      & 7(5/0/5)      & 1(0/0/0)            \\
UPX {v2.90}     & 5K        & 62K     & 11  & 1  & 0                      & 4(2/0/2)      & 0(0/0/0)            \\
\hline
\end{tabular}
%}

{\footnotesize
\smallskip
\begin{minipage}{0.6\linewidth}
opaque predicates: $k=16$ \--  Unk: query return Unknown \-- OP: proved as opaque  

stack tampering:  OK: \#\ret runtime behavior is genuine \-- KO: \#\ret violated at runtime

. a: proved aligned \-- d: proved disaligned \-- g: proved genuine  \-- s: proved to have a single target 
\end{minipage}
}

\end{table*}

\subsection{Other Discoveries}

\myparagraph{Opaque predicates}
Results revealed interesting patterns, for instance ACProtect tends to add opaque
predicates by chaining conditional jumps that are mutually exclusive like:
\texttt{jl  0x100404c ; jge  0x100404c}. In this example the second jump is
necessarily opaque since the first jump strengthens the path predicate, enforcing
the value to be lower. This example shows that our approach can detect
both invariant and contextual opaque predicates, and should also detect dynamic
opaque predicates since they are similar to contextual opaque predicates. Many
other variants of this pattern were found: \texttt{jp/jnp}, \texttt{jo/jno}, etc. 
Similarly, the well-known opaque predicate pattern \texttt{xor ecx, ecx; jnz} was
detected in \textsc{Armadillo}. As a value \texttt{xor}(ed) by itself always
return 0, the \texttt{jnz} is never taken.

\begin{figure}[htbp]
	\centering \includegraphics[width=0.8\columnwidth]{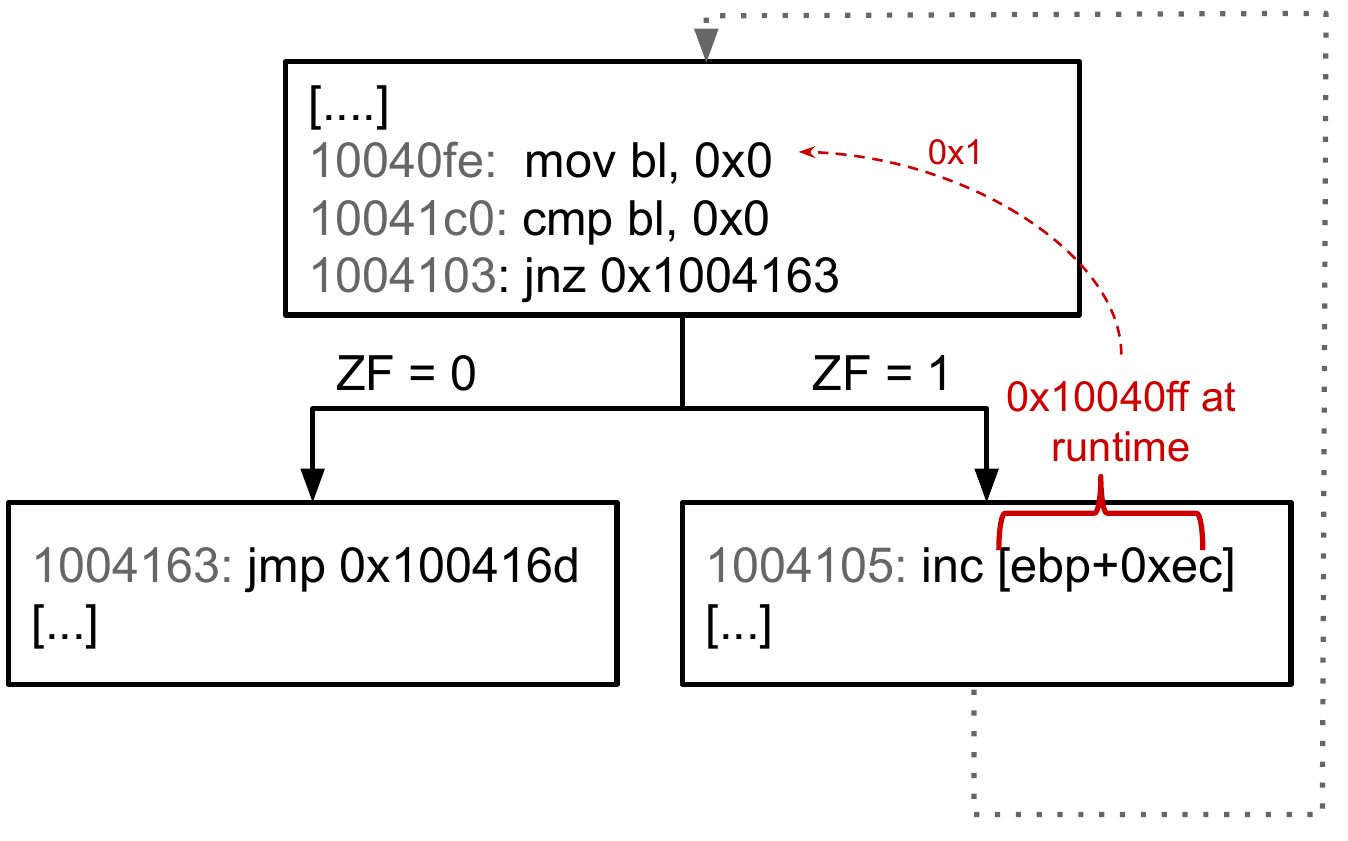}
	\caption{\aspack opaque predicate decoy}\label{fig:aspack_decoy}
\end{figure}

The dynamic aspect of \bbdse allowed to bypass some tricks that would misled a
reverser into flagging a predicate as opaque. A good example is a predicate found
in \aspack seemingly opaque but that turned not to be opaque due to a self-modification
(Figure. \ref{fig:aspack_decoy}). Statically, the predicate is opaque since \textsc{bl}
is necessarily 0 but it turns out that the second opcode bytes of the \textsc{mov bl, 0x0}
is being patched to 1 in one branch in order to take the other branch when looping back later on.

\myparagraph{Call/stack tampering}
From the call/stack tampering perspective and according to the taxonomy defined
in Section \ref{sec:encoding},  many different kinds of violations were detected. The first two
patterns found in ACProtect shown in Figures \ref{fig:acprotect_tampering1} and
\ref{fig:acprotect_tampering2} are respectively detected as \violated, \single,
\alignedd and \violated, \single, \disalignedd. Figures \ref{fig:aspack_tampering1},
\ref{fig:aspack_tampering2} and \ref{fig:aspack_tampering3} show 
three different kinds of violation found in \aspack. In the first example
(cf.~Figure~\ref{fig:aspack_tampering1}) the tampering is detected with labels
\violated, \disalignedd since the stack pointer read the \ret address at the wrong
offset. In the second example (cf.~Figure~\ref{fig:aspack_tampering2}), the return
value is modified in place. The tampering is detected with the \violated, \alignedd,
\single tags. The last example (cf.~Figure~\ref{fig:aspack_tampering3}), takes place
between the transition of two self-modification layers and the ret is used for 
 tail-transitioning to the packer  payload (i.e., the original unpacked program). This violation is detected
with \violated, \disalignedd, \single since the  analysis matches a \call far
upper in the trace which is disaligned. Note that
instruction {\tt push 0x10011d7} at address 10043ba is originally a {\tt push 0}, but it is patched by
instruction at address 10043a9, triggering the entrance in a new auto-modification layer
when executing it. This pattern reflects a broader phenomenon found in many packers
like {\tt nPack}, {\tt TELock} or {\tt Upack} having a single \ret tampered: these 
 packers perform their tail transition to the entrypoint
of the original (packed) program with {\tt push; ret}. Thus, such analysis allows to find precisely
that moment in the execution trace, where the payload is highly likely decompressed
in memory. 

\begin{figure}[htbp]
	\resizebox{\columnwidth}{!}{  \tiny
		\centering
		%\resizebox{\columnwidth}{!}{
		\def\arraystretch{1.2}
		\begin{tabular}{lll}
			address & mnemonic & comment \\
			\hline
			1004328 & call 0x1004318 & //push 0x100432d as return \\
			1004318 & add  [esp], 9  & //tamper the value in place \\
			100431c & ret            & //return to 0x1004n336 \\
		\end{tabular}
	}
	\caption{ACProtect violation 1/2}\label{fig:acprotect_tampering1}
\end{figure}

\begin{figure}[htbp]
	\resizebox{\columnwidth}{!}{ \tiny
		\centering
		%\resizebox{\columnwidth}{!}{
		\def\arraystretch{1.2}
		\begin{tabular}{lll}
			address & mnemonic & comment \\
			\hline
			1001000 & push 0x1004000 & \\
			1001005 & push 0x100100b & \\
			100100a & ret            & jump on the ret below \\
			100100b & ret            & jump on 0x1004000 \\
		\end{tabular}
	}
	\caption{ACProtect violation 2/2}\label{fig:acprotect_tampering2}
\end{figure}

\begin{figure}[htbp]
	\resizebox{\columnwidth}{!}{  \tiny
		\centering
		%\resizebox{\columnwidth}{!}{
		\def\arraystretch{1.2}
		\begin{tabular}{llll}
			address & len & mnemonic & comment \\
			\hline
			1004a3a & 5 & call  0x1004c96 & //push 0x1004a3f as return site \\
			1004c96 & 5 & call  0x1004c9c & //push 0x1004c9b as return site\\
			1004c9c & 1 & pop esi         & //pop return address in esi \\
			1004c9d & 5 & sub esi, 4474311& \\  
			1004ca3 & 1 & ret             & //return to 0x1004a3f \\
		\end{tabular}
	}
	\caption{ASPack violation 1/3}\label{fig:aspack_tampering2}
\end{figure}

\begin{figure}[htbp]
	\resizebox{\columnwidth}{!}{  \tiny
		\centering
		%\resizebox{\columnwidth}{!}{
		\def\arraystretch{1.2}
		\begin{tabular}{lll}
			address & mnemonic & comment \\
			\hline
			1004002 & call 0x100400a & //push 0x1004007 as return \\
			1004007 & .byte{invalid} & //invalid byte (cannot disassemble) \\
			1004008 & [...]          & //not disassembled \\
			100400a & pop ebp        & //pop return address in ebp  \\                        
			100400b & inc ebp        & //increment ebp \\
			100400c & push ebp       & //push back the value \\
			100400d & ret            & //jump on 0x1004008\\
			%1004008 eb 04          jmp 0x100400e
		\end{tabular}
	}
	\caption{ASPack violation 2/3}\label{fig:aspack_tampering1}
\end{figure}

\begin{figure}[htbp]
	\centering
	\resizebox{\columnwidth}{!}{ \tiny
		\def\arraystretch{1.2}
		\begin{tabular}{llcl}
			address & mnemonic & layer & comment \\
			\hline
			10043a9 & mov [ebp+0x3a8], eax & 0 & //Patch push value at 10043ba* \\
			10043af & popa           & 0 & //restore initial program context \\
			10043b0 & jnz  0x10043ba & 0 & //enter last SM layer (payload) \\
			\hline
			& Enter SMC Layer 1  &   & \\
			\hline
			10043ba & push 0x10011d7 & 1 & //push the address of the entrypoint \\
			10043bf & ret            & 0 & //use ret to jump on it \\
			10011d7 & [...]          & 1 & //start executing payload\\
		\end{tabular}
	}
	\smallskip
	{\scriptsize *(at runtime eax=10011d7 and ebp+0x3a8=10043bb)}
	
	\caption{ASPack violation 3/3}\label{fig:aspack_tampering3}
\end{figure}

\subsection{Conclusion}

By detecting   opaque predicates and call/stack tampering on packers with multi-million trace length, 
this experiment clearly demonstrates both   the ability of \bbdse  to scale to realistic  obfuscated examples (without any prior-knowledge  of the protection schemes) and its usefulness.   
%  
%
%
%Thus, the \bbdse scales to these trace lengths and allowed to find real-cases
%without prior-knownledge of inner working of packers. 
%
This study yields also a few unexpected and valuable  insights on the inner working on the considered packers, such as some kinds of protections 
or the location of the jump to the entrypoint of the original unpacked program.

%============================ XTUNNEL ==========================
%===============================================================
\section{Real-world malware: X-Tunnel} \label{sec:xtunnel}

\subsection{Context \& Goal}

\myparagraph{Context} As an application of the previous techniques we focus in this section on the heavily
obfuscated \xtunnel malware.
\xtunnel is a ciphering proxy component allowing the \xagent malware to reach the
command and control (CC) if it cannot reach it directly~\cite{calvet_visiting_2016}.
It is usually the case for machines not connected to internet but reachable from
an internal network. These two malwares are being used as part of target attack
campaigns (APT) from the APT28 group also known as Sednit, Fancy Bear, Sofacy or
Pawn Storm.
This group, active since 2006, targets geopolitical entities and is supposedly highly
tight to Russian foreign intelligence. Among alleged attacks, noteworthy targets are
NATO~\cite{trend_micro_operation_2014}, EU institutions~\cite{eset_research_sednit_2015},
the White House~\cite{trend_micro_operation_2015}, the German
parliaments~\cite{von_gastbeitrag_digital_2015} and more recently the American Democrate National Comittee
DNC~\cite{alperovitch_bears_2016} that affected the running of elections. This group also makes use of many
0-days~\cite{mehta_cve-2016-7855:_2016} in Windows, Flash, Office, Java and  also operate other malwares
like rootkits, bootkits, droppers, Mac 0SX malwares~\cite{creus_sofacys_2016} as
part of its ecosystem.

\myparagraph{Goal}
This use-case is based on 3 \xtunnel samples\footnote{We warmly thank Joan Calvet
for providing the samples.} covering a 5 month period (if timestamps are correct).
While Sample \#0 is not obfuscated and can be straightforwardly analyzed, Samples
\#1 and \#2 are, and they are also much larger than Sample \#0 (cf.~Table
\ref{tab:xtunnel_infos}). The main issue raised here is: 

% 
% -- a quick manual inspection reveals  

%    The difference is that the first sample is not obfuscated
%while the two others are. If timestamp are
%corrects the samples cover a 5 month period of development. The main issue raised is:

\begin{center}
	G1: \textit{Is there new functionalities in  the obfuscated samples?}
\end{center}

Answering this question requires first to be able to analyse the obfuscated binaries. Hence we focus here on a second goal:

\begin{center}
	G2: \textit{Recover a de-obfuscated version of the obfuscated samples.}
\end{center}

%% Table \ref{tab:xtunnel_infos} gives the basic information about the samples where
%% the two obfuscated are twice bigger in instruction than the sample \#0.

\begin{table}[htpb]
	\caption{Samples infos}\label{tab:xtunnel_infos}
	\centering
	\resizebox{\columnwidth}{!}{
	\def\arraystretch{1.2}
	\begin{tabular}{|l|l|l|l|}
		\hline
		& Sample \#0  & Sample \#1  & Sample \#2 \\
		& \footnotesize 42DEE3[...] & \footnotesize C637E0[...] & \footnotesize 99B454[...] \\
		\hline 
		\hline
		obfuscated     & No         & Yes        & Yes \\
		\hline
		size		   & 1.1 Mo     & 2.1 Mo     & 1.8 Mo \\
		\hline
		creation date  & 25/06/2015 & 02/07/2015 & 02/11/2015 \\
		\hline
		\#functions    & 3039       & 3775       & 3488 \\
		\hline
		\#instructions & 231907     & 505008     & 434143 \\
		\hline
	\end{tabular}
	}
\end{table}

We show in the latter how \bbdse can solve goal G2, and we give hints on what is to be done to  solve G1.

\myparagraph{Analysis context} Obfuscated samples appeared to contain a tremendous amount of opaque predicates. 
As a consequence, our goal is to detect and remove all opaque predicates in order to
remove the dead-code and meaningless instructions to hopefully obtain a de-obfuscated CFG.
This deobfuscation step is a prerequisite for later new functionality finding.

The analysis here has to be performed statically:
\begin{itemize}
	\item as the malware is a network component, it requires to connect
	to the CC server, which is not desirable; 
	\item following the same line, many  branching conditions are network-event based, thus unreliable and more hardly
	reproducible (and  would also require infected clients for connection to \xtunnel);
	\item \xtunnel does not look to use any self-modification obfuscation or
	neatly tricks to hamper the disassembly. Thus the whole disassembled code
	is available.
\end{itemize}
The only  difference with previous experiments is the need to test the two branches for each conditional jumps.

\subsection{Analysis}

\myparagraph{OP detection} 
The analysis performs  a \bbdse on every conditional jumps of the
program, testing systematically both branches. Taking advantage of previous experiments, 
we set the the bound $k$ to 16.  The solver used is \ZE with
a 6s timeout. If both branches are \unsat,  the predicate is considered dead,  as the
unsatisfiability is necessarily due to path constraints indicating that the predicate
is not reachable. 

\myparagraph{Code simplification} We perform three additional computations in complement
to the opaque predicate detection:

\begin{itemize}
	\item{\bf predicate synthesis} recovers the high-level predicate of an opaque
	predicate by backtracking on its logical operations. The goal of this analysis
	is twofold: (1) indexing the different kind of predicates used and (2) identifying
	instruction involved in the computation of an OP denoted {\it spurious instructions}
	(in order to remove them);
	\item{\bf liveness propagation} %standard liveness propagation 
        based on obfuscation-related 
	data aims at marking instruction by theirs status, namely {\it alive, dead, spurious}; 

	\item{\bf reduced CFG extraction} extracts the de-obfuscated CFG based on the liveness
	analysis. 
\end{itemize}

\subsection{Results}

\myparagraph{Execution time} Table \ref{tab:xtunnel_exectime} reports the execution
time of the the \bbdse and predicate synthesis. The predicate synthesis takes a
non-negligible amount of time, yet it is still very affordable, and moreover our implementation is far from optimal. 
%The 10 predicates per second includes  sub-path creation, solving and communications
%the engine (\binsecse) and the IDA plugin (\idasec) via network communications.

\begin{table}[htpb]
	\caption{Execution time}\label{tab:xtunnel_exectime}
	\centering
	\def\arraystretch{1.2}
		\resizebox{\columnwidth}{!}{
	\begin{tabular}{|l|l|l|l|l|}
		\hline
		          & \#preds  & DSE        & Synthesis  & Total      \\% & pred/sec\\
		\hline 
		\hline
		Sample \#1 & 34505    & 57m36      & 48m33      & 1h46m      \\% & $\sim$10 \\
		%&          & (3456.00s) & (2913.22s) & (6369.29s) &          \\
		\hline
		Sample \#2 & 30147    & 50m59      & 40m54      & 1h31m      \\% & $\sim$10 \\
		%&          & (3059.80s) & (2454.9s)  & (5514.70s) &          \\
		\hline
	\end{tabular}
	}
\end{table}

\myparagraph{OP diversity}
Each sample presents a very low diversity of opaque predicates. Indeed, solely
$7x^2 - 1 \neq x^2$ and $ \frac{2}{x^2 + 1} \neq y^2 + 3$ were found. Table
\ref{tab:xtunnel_po} sums up the distribution of the different predicates.
The amount of predicates and their distribution supports the idea that they
were inserted automatically and picked randomly.

\begin{table}[htpb]
	\caption{Opaque predicates variety}\label{tab:xtunnel_po}
	\centering
	\def\arraystretch{1.3}
		\resizebox{0.8\columnwidth}{!}{
	\begin{tabular}{|l|c|c|}
		\hline
		%& \bf OP \#1       & \bf OP \#2                    \\
		& $7y^2 - 1 \neq x^2$ & $ \frac{2}{x^2 + 1} \neq y^2 + 3$\\
		\hline 
		\hline
		Sample \#1 & 6016 (49.02\%)   & 6257 (50.98\%) \\
		\hline
		Sample \#2 & 4618 (45.37\%)   & 5560 (54.62\%) \\
		\hline
	\end{tabular}
		}
\end{table}

\begin{figure}[h!tbp]
	\centering
	%\resizebox{\textwidth}{!}{
	%\begin{mdframed}
	% === rtn0 ====
	\begin{subfigure}{.49\columnwidth}
		\centering
		\includegraphics[width=0.65\linewidth]{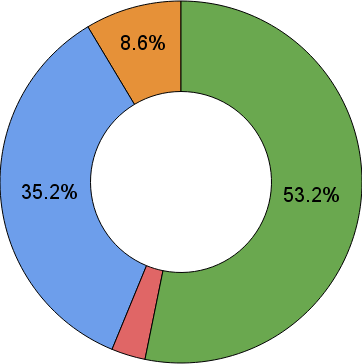}
		\caption{\small OP results Sample \#1}
		\label{fig:false_pos}
	\end{subfigure}
	\begin{subfigure}{.49\columnwidth}
		\centering
		\includegraphics[width=0.65\linewidth]{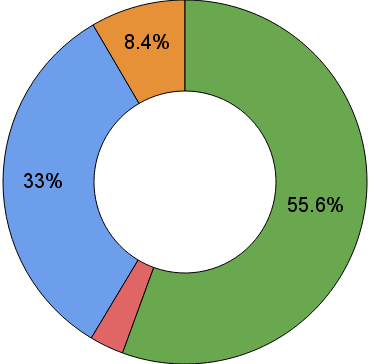}
		\caption{\small OP results Sample \#2}
		\label{fig:likely}
	\end{subfigure}
	%\end{mdframed}
	%}
	
	\smallskip {\footnotesize \myred{$\blacksquare$} FN \mygreen{$\blacksquare$} OK \myblue{$\blacksquare$} Opaque \myorange{$\blacksquare$} FP}
	\caption{Graph of opacity distribution}\label{fig:xtunnel_opacity}
\end{figure}

\myparagraph{Detection results}
As the diversity of opaque predicates is very low, we are able to determine, with
quite a good precision, the amount of false negatives and false positives based
on the predicate synthesized. If a predicates matches one of the two OP and was
detected OK, then we considered it false negative (respectively false positive).
Results are given in Table \ref{tab:xtunnel_results} and Figure~\ref{fig:xtunnel_opacity}.
The detection rate is 
satisfactory as false negatives only represent 3\% of all predicates.  Conversely,  
8.4 to 8.6\% of false positive are wrongly tagged opaque. 

%We only test one path
%for each predicates. As consequence, most false positives are due the wrong path
%selection. Figure given in annex \ref{fig:false_pos} shows an example where the
%path selected go through the {\tt xor ebx, ebx} which makes the predicate
%{\tt cmp ebx, 0x7f} to be always false due to the wrong path selection (always
%the shortest from the function entry).

\begin{table}[htpb]
	\caption{Opaque predicates evaluation}\label{tab:xtunnel_results}
	\centering
	\def\arraystretch{1.3}
	\resizebox{\columnwidth}{!}{
	\begin{tabular}{|l|l|l|l|l|l|l|l|}
		\hline
		& \multirow{2}{*}{\#pred} & \multicolumn{2}{c|}{OK}  & \multicolumn{2}{c|}{OP} & \multicolumn{2}{c|}{Likely} \\
		\cline{3-8}
		&                         & OK        & FN           & OP       & FP           & OK      & OP  \\
		\hline 
		\hline
		Sample \#1& 34505   & 17197   & 1046    & 11973    & 2968    & 1156    & 165   \\
		&         & (49.8\%)& (3.0\%) & (34.7\%) & (8.6\%) & (3.4\%) & (0.4\%) \\
		\hline
		Sample \#2& 30147   & 16148    & 914      & 9790    & 2543    & 606     & 146 \\
		&         & (53.7\%) & (3.0\%) & (32.5\%) & (8.4\%) & (2.0\%) & (0.5\%) \\
		\hline
	\end{tabular}
	}
	{\footnotesize likely: predicates were both branches were \unsat}
\end{table}

\myparagraph{Dependency evaluation}
As seen previously, a large $k$ bound can lead to false positive due to nested
opaque predicates while in the meantime a low bound misses some predicates.
Finding the right balance is still an important issue, but results with 12138 OP
detected against 1046 false negative tend to confirm that such a low bound
is a good trade-off. Across the two samples, the maximum distance between a predicate
and its variable definition where 230 (Sample \#1) and 148 (Sample \#2). Still, the average
computed on all the OPs yield an average of 8.7.

\myparagraph{Difference with \ollvm} Interesting differences with OP found in \ollvm are to
be emphasized. Firstly, there 
is more interleaving between the payload and the OPs computation. Some  meaningful instructions are often encountered within the 
predicate computation. Secondly, while \ollvm OPs are really local to the basic
block, there are here some code sharing between predicates.
As a consequence, predicates are not fully independent from one another. Also,
the obfuscator uses local function variables to store temporary 
results at the beginning of the function for later usage in opaque predicates.
This leads to   increase  the depth of the dependency chain and
to complicate the detection.

\begin{figure*}[!thbp]
	\centering
	%\resizebox{\textwidth}{!}{
	% === rtn4 ====
	\begin{subfigure}{.33\textwidth}
		\centering
		\includegraphics[width=0.7\linewidth,height=20em]{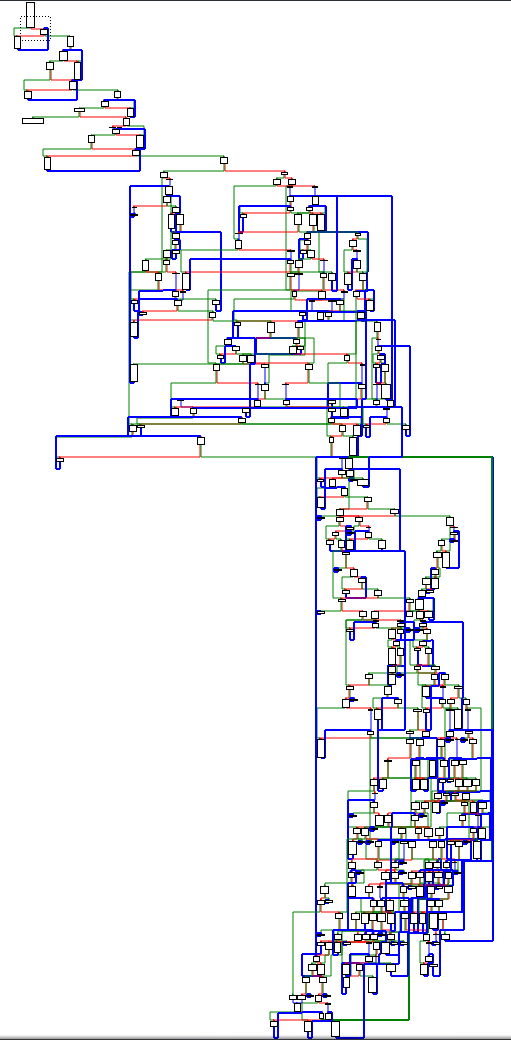}
		\caption{Original function CFG}
		\label{fig:rtn4_cfg}
	\end{subfigure}%
	\begin{subfigure}{.33\textwidth}
		\centering
		\includegraphics[width=0.7\linewidth,height=20em]{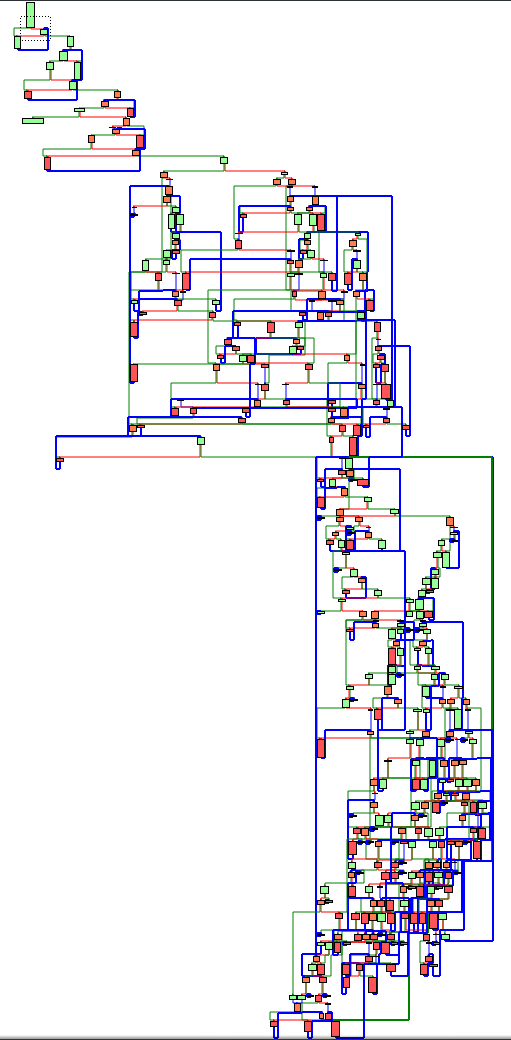}
		\caption{CFG tagged}
		\label{fig:rtn4_color}
	\end{subfigure}
	\begin{subfigure}{.33\textwidth}
		\centering
		\includegraphics[width=0.7\linewidth,height=20em]{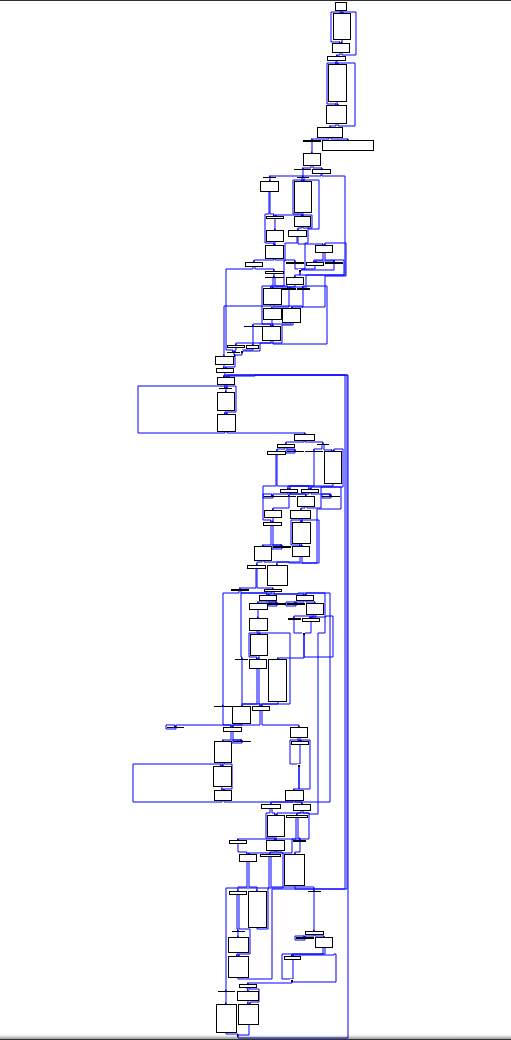}
		\caption{CFG extracted}
		\label{fig:rtn4_extract}
	\end{subfigure}
	%}
\caption{Examples of CFG extraction}\label{fig:summary_extraction}
\end{figure*}

\myparagraph{Code simplification, Reduced CFG extraction}
Table \ref{tab:xtunnel_simplification}
shows the number of instructions re-classified based on their status. The dead
code represents 1/4 of all program instructions. Computing the difference with the
original non-obfuscated program shows a very low difference. Therefore,
the simplification pass allowed to retrieve a program which is roughly the size of
the original one. The difference is highly likely to be due to the false negatives
or missed {\it spurious} instructions. % 
Finally, Figure~\ref{fig:summary_extraction} shows a function originally (a), with the status tags (b),
and the result after extraction (c) using tags (red:dead, orange:spurious, green:alive). Although
the CFG extracted still containing noise, it allows a far better understanding of the function
behavior. A demo video showing the deobfuscation of a \xtunnel function with
\binsec and \idasec is available as material for this paper\footnote{https://youtu.be/Z14ab\_rzjfA}.

\begin{table}[!htpb]
	\caption{Code simplification results}\label{tab:xtunnel_simplification}
	\centering
	\def\arraystretch{1.3}
	\resizebox{\columnwidth}{!}{
	\begin{tabular}{|l|c|c|c|c|c|}
		\hline
		& \#instr   & \#alive  & \#dead  & \#spurious  & diff sample \#0$^\dagger$ \\
		%	&           &          &         &             & (231,907 instrs)  \\ 
		\hline 
		\hline
		\multirow{2}{*}{Sample \#1} & \multirow{2}{*}{507,206}   & 279,483  & 121,794 & 103,731     & \bf\multirow{2}{*}{47,576} \\
		&                            & (55\%)   & (24\%)  & (20\%)      &        \\
		\hline
		\multirow{2}{*}{Sample \#2} & \multirow{2}{*}{436,598}   & 241,177  & 113.764 & 79,202      & \bf\multirow{2}{*}{9,270}  \\
		&                            & (55\%)   & (26\%)  & (18\%)      &        \\
		\hline
	\end{tabular}
	}
	
	\smallskip
	$\dagger$ Sample \#0: 231,907 instrs
\end{table}

%\myparagraph{Reduced CFG extraction}

\subsection{Conclusion}

\myparagraph{About the case-study}
We have been able to automatically detect  opaque predicates in the two obfuscated samples of the \xtunnel malware, leading a significant (and automatic) simplification 
of these codes -- removing all  spurious and dead instructions.  Moreover, we have gained insights (both strengths and weaknesses) into the inner working of \xtunnel protections. 
Hence, we consider that goal G2 has been largely achieved.  
%from the obfuscated samples of 
% 
%
%This case-study showed how to extract a readable CFG version of them.
In order to  answer to the initial question (G1), some similarity algorithms should now 
be computed between the non-obfuscated and simplified samples, in order to detect if some new functions have been added to the code. Moreover, 
our analysis also  pinpoints the protected functions (a small minority), and this information can surely be taken into account.    For now,  this second 
analysis step is left as a future work.

\myparagraph{About \xtunnel protections}
The obfuscation found here are quite sophisticated compared with existing
opaque predicates found in the state-of-the-art. It successfully manages
to spread the data dependency across a function so that some predicates
cannot be solved locally at the basic block level. Hopefully, this is not
a general practice across predicates so that the \bbdse works very well in the
general case. The main issue of the obfuscation is the low diversity of
opaque predicates in the way that some pattern matching can come in relay
of symbolic approaches to classify a posteriori false positives and false
negatives.

%% \myparagraph{Possible improvements}
%% For each conditional jumps, computing more k-paths would reduce the risk of having
%% selected the wrong one. 
%% Moreover, performing a taint on data would enable to detect missing data
%% dependencies while keeping the scalability of the Backward-Bounded approach
%% without being hampered by path constraints making formulas \unsat for both branches.

%====================== SPARSE DISASSEMBLY =====================
%===============================================================
\section{Application: Sparse Disassembly} \label{sec:sparse}

%\todo[inline]{SB: make sure all points claimed in intro (tables) is explained here}

\subsection{Principles}

%% --------------- SB : bon texte, mais redondant 
%%
%% Main algorithms algorithms for disassembly are linear-sweep and recursive
%% traversal. Linear-sweep just iterate every code sections and linearly decode
%% every bytes assuming it is code. This technic is resilient to some binary
%% constructs like indirect jumps but face the risk to over-disassemble the program
%% since some bytes are just padding or data. Recursive traversal follow the
%% control flow of the program while disassembling. It ensure a safe disassembly
%% of the program (does not decode data instruction) but it is particularly 
%% vulnerable to indirect/calculated jumps. Thus without heuristics recursive
%% traversal is unable to know where instructions like jump eax or ret are going
%% to jump since the value is computed at runtime. Moreover, recursive travesal
%% usually assume that a \ret, return to its caller. Thus, when reaching a \call the
%% recursive traversal algorithm will disassemble both the call target and the 
%% instruction following the \call (usual return site). This assumptions hold
%% in well-formed programs but can be fooled easily by obfuscators. Indeed, with
%% stack tampering a \ret does not necessarily returns to its caller and a static
%% disassembler as no way to know that a branch in a program is dead.

As already explained, static  and dynamic disassembly methods tend to have complementary strengths and weaknesses, and  \bbdse 
is the only robust approach targeting infeasibility questions. 
Hence, we propose  {\it sparse disassembly}, an algorithm based on recursive disasssembly
reinforced with a dynamic trace and complementary information about obfuscation 
(computed by \bbdse) in order to provide a more precise disassembly of obfuscated codes.  
The basic idea is to  {\it enlarge} and initial  dynamic disassembly by a cheap syntactic disassembly  {\it in a guaranteed way}, following 
information from \bbdse, hence 
getting the best of dynamic and static approaches.

\begin{figure}[htbp]
	\centering
	\includegraphics[width=0.6\columnwidth]{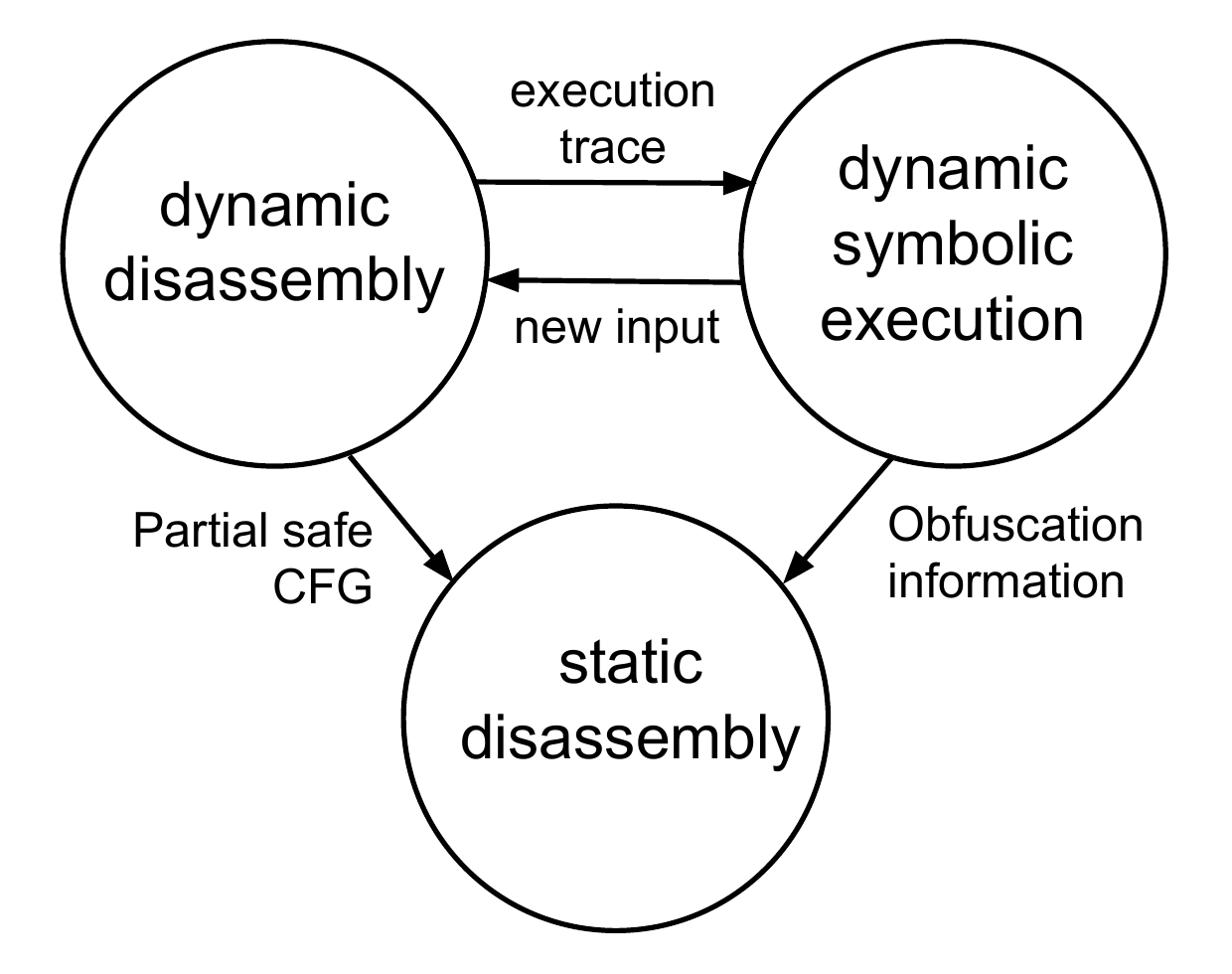}
	\caption{Sparse disassembly combination}\label{fig:combination_schema}
\end{figure}

The approach takes 
advantage of the two analyses presented in Sections \ref{sec:po} and \ref{sec:callret} in the following way (cf.~Figure\ref{fig:combination_schema}): 
\begin{itemize}

	\item use dynamic values found in the trace to keep disassembling after indirect
	jump instructions; 

	\item use opaque predicates found by \bbdse to avoid disassembling dead branches 
	(thus limiting the number of recovered non legit instructions); 

	\item use stack tampering information found by \bbdse to disassemble \ret targets in case
	of violation, as well as not to disassemble the return site of the \call in this case. 
\end{itemize}

\myparagraph{Implementation} 
A preliminary  version of this algorithm has been integrated in \binsec, taking advantage of the existing recursive  disassembly
algorithm. The \bbdse procedure sends OP and \ret\ information to the modified recursive disassembler, which takes the information into account.   
%  hence perform an interaction between the disassembly
%engine and the dynamic symbolic execution engine. \mynote{quoi dire de plus ?}

\subsection{Preliminary Evaluation}

%% 1ere xp : precision, controlled xp
%%    sparse vs : ida, objdump, rec tout court
%%    po  (ollvm)
%%    cs  (tigress)   

%%    ground truth, report exact covering  
%%    // nb of instr good approx : op : less is better, tigress : it happens that here it is also the case

%%    conclusion : on est nickel (pas le cas de autres)
%%                 difference notable vs ida

%% 2e xp :  larger examples, regarde aussi capacité à élargir la trace
%%   sparse vs : aussi trace
%%   op only

We report two sets of experiments, designed to assess the precision of the approach and its ability to enlarge an initial dynamic trace. 
We compare our method mainly to the well-known disassembly tools \IDA and \objdump. \IDA relies on a combination of recursive disassembly, linear sweep and dedicated heuristics.  
\objdump performs only liner sweep.

\myparagraph{Precision} In the first evaluation, we compare these different tools on simple programs obfuscated either by O-LLVM (opaque predicates) or Tigress (stack tampering). 
In each experiment, we compare the set of disassembled instructions with the set of legitimate instructions of the obfuscated program (i.e.~those instructions which can be part of a real execution). 
It turns out on these small examples that all methods are able to find all  the legitimate instructions, yet they may nor may not be lured into dead instructions introduced by 
obfuscation. 

Tables  \ref{tab:sparse_po} and \ref{tab:sparse_callret} present our results. We report for each program and each disassembly method the number of recovered instructions. 
It turns out that this information is  representative of  the quality of the disassembly (the less instruction, the better), given the considered obfuscations and 
 the fact that here all methods recover all legitimate instructions (actually, all results have been checked manually). % and that the classes of obfuscation considered and 

\begin{table}[htbp]
\caption{Sparse disassembly opaque predicates}\label{tab:sparse_po}
\centering
\resizebox{\columnwidth}{!}{
\def\arraystretch{1.2}
\begin{tabular}{|l|c|c||c|c|c|c|}
\hline
\multirow{3}{*}{sample} &  & \multicolumn{4}{c|}{Obfuscated} & gain \\
\cline{3-6}
            & no             & \multirow{2}{*}{perfect}  &  \multirow{2}{*}{\IDA} & \multirow{2}{*}{\objdump} & \multicolumn{1}{c|}{\ \binsec\ }  & vs \IDA \\
            & obf.                &          &      &         &    {\bf sparse} &  (sparse)          \\
\hline
simple-if   & 37 & \bf 185    & 240   & 244 &  \bf 185   & 23,23\% \\
\hline
huffman     & 558 & \bf 3226  & 3594  & 3602 &  \bf 3226 & 10,26\% \\
\hline
mat\_mult   & 249 & \bf 854   & 1075  & 1080 &  \bf 854  & 20,67\% \\
%multiply1   &     &          &      &      &                 
\hline
bin\_search & 105 & \bf 833   &  1110 & 1115 &  \bf 833  & 24,95\% \\
\hline
bubble\_sort& 121 & \bf 1026  & 1531  & 1537 &  \bf 1026 & 32,98\% \\
\hline
\end{tabular}
}
\end{table}

\begin{table}[htbp]
\caption{Sparse disassembly stack tampering}\label{tab:sparse_callret}
\centering
\resizebox{\columnwidth}{!}{
\def\arraystretch{1.2}
\begin{tabular}{|l|c|c|c|c|c|c|}
\hline
\multirow{3}{*}{sample} &  & \multicolumn{4}{c|}{Obfuscated} & gain \\
\cline{3-6}
            & no        & \multirow{2}{*}{perfect} &  \multirow{2}{*}{\IDA} & \multirow{2}{*}{\objdump} & \multicolumn{1}{c|}{\ \binsec\ }  & vs \IDA \\
            & obf.      &          &      &          & {\bf sparse} &  (sparse)          \\
\hline
simple-if  & 37 & \bf 83 & 95 & 98 &  \bf 83 & 14.45\% \\
\hline
huffman    & 558 & \bf 659 & 678 & 683 &  \bf 659 & 2.80\% \\
\hline
mat\_mult  & 249 & \bf 461 & 524 & 533 &  \bf 461 & 12.0\% \\
\hline
bin\_search & 105 & \bf 207 & 231 & 238 & \bf 207 & 10.39\% \\
\hline
bubble\_sort & 121 & \bf 170 &  182 & 185 &  \bf 170 & 6.6\% \\
\hline
\end{tabular}
}
\end{table}

In both cases, sparse disassembly achieves a {\it perfect score} -- recovering all but only legitimate instructions, performing better than \IDA and \objdump. 
Especially, when  opaque predicates are considered, sparse disassembly recovers up to 32\% less instructions than \IDA.

\myparagraph{Improvement over dynamic analysis} We now seek to assess whether sparse disassembly can indeed enlarge 
a dynamic analysis in a significant yet guaranteed way, i.e.~without adding dead instructions. We consider 5 larger \coreutils programs   obfuscated with O-LLVM. %programs 
We compare sparse disassembly to dynamic analysis (starting from the same trace). 

Here again, the number of recovered instructions is a good metric of precision (the bigger, the better), since 
both methods {\it report only legitimate instructions} on these examples (we checked that \bbdse was able to find all inserted opaque predicates). 
Results are reported in Table \ref{tab:sparse_coreutils}. We also report the output of \IDA and \objdump in order to give an upper-bound  of 
the number of instructions, yet the two tools recover many dead instructions. 

\begin{table}[htbp]
\caption{Sparse disassembly \coreutils}\label{tab:sparse_coreutils}
\centering
\resizebox{\columnwidth}{!}{
\def\arraystretch{1.2}
\begin{tabular}{|l|c|c|c||c|c|}
\hline
\multirow{3}{*}{sample} &  & \multicolumn{4}{c|}{Obfuscated} \\
\cline{3-6}
            &    Tr.len   & \multirow{2}{*}{\objdump} & \multirow{2}{*}{\IDA}   & Dynamic &  {\ \binsec\ } \\
            &             &         &       &  disas. &   {\bf sparse} \\
\hline
basename    & 1,783        & 20,776   & 20,507 & \bf 1,159    & \bf 7,894  \\
\hline
env         & 3,692        & 19,714   & 19,460 & \bf 477     & \bf 6,743  \\
\hline
head        & 17,682       & 32,840   & 32,406 & \bf 1,299   & \bf 19,807 \\
\hline
mkdir       & 1,436        & 57,238   & 56,767 & \bf 1,407   & \bf 10,428 \\
\hline
mv          & 14,346       & 115,278  & 114,067& \bf 5,261   & \bf 81,596\\
\hline
\end{tabular}
}
\end{table}

 Actually, these experiments demonstrate that sparse disassembly is an effective way to {\it enlarge a dynamic disassembly}, 
 in a both {\it significant and guaranteed manner}. Indeed, sparse disassembly recovers between 6x and 16x more instructions than dynamic disassembly, yet 
 it still recovers  much less than linear sweep -- due to the focused approach of dynamic disassembly and the guidance of \bbdse. 
 Hence, sparse disassembly stays close to the original trace.

\myparagraph{Conclusion} The carried experiments showed very good and accurate results on controlled samples, achieving 
 perfect disassembly. From this stand-point, sparse disassembly performs better than combination of both recursive
and linear like in \IDA, with up to 30\% less recovered instructions than \IDA. The \coreutils experiments showed that sparse disassembly 
 is also an effective way to enlarge a dynamic disassembly  
in a both significant and guaranteed manner. 
In the end, this is a clear demonstration of infeasibility-based information used in the context of disassembly.
  
Yet, our sparse disassembly algorithm is still very preliminary. It is currently  limited by the inherent weaknesses of recursive disassembly (rather than sparse disassembly shortcomings), for example 
the handling of computed jumps would require advanced pattern techniques.   

%This shows a great application of how the infeasibility-based queries and especially OPs and
%CST can be used in the context of disassembly.
%Thus, 
%sparse disassembly performs better on obfuscated samples than recursive disassembly alone and limitate the 
%over-disassembly shortcoming of linear-sweep based approaches.  

%\todo[inline]{SB: assez negatif je trouve, alors que le but est d'étendre une trace de manière safe > c'est atteint !!}

%======================= discussion =======================
%============================================================
\section{Discussion: Security Analysis}  \label{sec:discussion}

From the attacker point of view, two main counter-measures can be employed to hinder
our approach. We present them as well as some possible mitigation.  

%% \mypar The first counter-measure  is to artificially spread the computation  of the obfuscation scheme 
%% over long sequence of codes
%% usage of logical values or increasing the number of paths leading to the property
%% to prove. It would force to increase the bound for the \bbdse or to increase the
%% number of paths to test which necessarily imply more computation time. Nevertheless,
%% it is often not necessary to backtrack all the dependencies to prove the predicate
%% opacity. An example is given in \xtunnel were many predicates have a dependency
%% chain longer than the bound (k=16) but this value was most of the time sufficient
%% to gather enough constraints. A mitigation for predicate with far dependencies is
%% to introduce a more generic notion of ``k'' bound which is not a strict number of
%% instructions but rather a def-use chain length or some formula complexity criterias.

\mypar The first counter-measure  is to artificially spread the computation  of the obfuscation scheme 
over a long sequence of code, 
%usage of logical values or increasing the number of paths leading to the property
%to prove. 
hoping either to evade the ``k'' bound of the analysis (false negatives) or to force a too high value for k (false positives or timeouts). 
% 
%It would force to increase the bound for the \bbdse or to increase the
%number of paths to test which necessarily imply more computation time. 
Nevertheless,
it is often not necessary to backtrack all the dependencies to prove infeasibility. 
An example is given in \xtunnel were many predicates have a dependency
chain longer than the chosen bound (k=16, chain up to 230) but this value was most of the time sufficient
to gather enough constraints to prove predicate opacity. 
Moreover, a  very good mitigation for these ``predicates with far dependencies'' is
to rely on  a more generic notion of the k bound,  based for example on  def-use chain length or some formula complexity criterias rather than a strict number of instructions.

%%\todo[inline]{on parle de augmenter le nb de chemins ? pas trop de très bonne réponse (ok si augmentation locale)}

% autre : augmenter le nombre de chemins 
%
%    roughly : def-use -> augmente la longueur (borne k)                   -> on manque des predicats opaques (faux negatifs, pas trop grave)
%              #path   -> augmente la largeur, taille du pre_k             -> on manque des chemins, donc on fait plus facilement des opaques (faux positifs, très grave)
%  
%    def-use : aborne + maline
% 
%    #path   : si local  : relancer disas (syntaxique, local) pour retrouver les branches, + merger et envoyer solveur (as veriT)
%                         voir local DSE, comme godefroid-cristakis (icse recent)
% 
%                                 -- mais both approaches gêner par computed jumps 
%
%
%              si global : semble très chaud ... là encore : ajouter paths de ida ? statique ? 
%
%              rmq : complexifie bcp quand même l'obfuscation   
%
%

\mypar The second counter-measure is to introduce hard-to-solve predicates (based for example on 
Mixed-Boolean Arithmetic~\cite{hutchison_information_2007} or cryptographic hashing
functions) in order to  lead
to inconclusive solver responses (timeout). As we cannot directly influence the
solving mechanism of SMT solvers, there is no clear mitigation from the defender perspective.   
Nonetheless,  solving such hard formula is an active topic research and some progress can be expected in a middle-term.  
Moreover,  triggering a timeout is already a valuable information, since \bbdse with reasonable k bound usually does not timeout. 
The defender can take advantage of it by manually inspecting the timeout root cause and deduce a hard-to-solve (in-)feasible pattern,  which can now  be 
detected through  mere syntactic matching.    
Finally,  such counter-measures would greatly complicate the malware design (and its cost!) and a
careless insertion of such complex patterns could lead to atypical code structures  
prone to relevant malware signatures.

%\todo[inline]{ajouter : sujet chaud de recherche ? driving challenge for smt-solvers ? jouer sur ``pb for any symb approach'' ? }

%======================= Related Work =======================
%============================================================
\section{Related Work} \label{sec:related}

\myparagraph{DSE and deobfuscation}
%
%\mynote{Sebastien DSE, et backward tu connais mieux que moi..}
%
Dynamic Symbolic Execution has been used in multiple situations
to address obfuscation, generally for discovering new paths in the code to analyze. Recently, Debray at al.~\cite{ccs2015,sp15}   % kruegel and co ?
used DSE against conditional and indirect jumps, VM and return-oriented 
programming on various packers and malware in order to prune the obfuscation
from the CFG. Mizuhito {\it et al.} also addressed exception-based obfuscation  using 
such techniques~\cite{Hai2016}. Recent work from Ming {\it et al.}~\cite{loop} used (forward) DSE %of BAP~\cite{bap}
to detect different classes of opaque predicates. Yet, their technique has difficulties to scale
due to the trace length (this is consistent with experiments in Section \ref{sec:vs-dse}). 
Indeed, by doing it in a forward manner they needlessly have to deal with the whole path predicate
for each predicate to check. As consequence they make use of taint to counterbalance
which far from being perfect brings additional problems (under-tainting/over-tainting).

DSE is designed to prove the reachability of certain parts of code (such as path, branches or instructions). It is complementary to \bbdse in that 
it addresses feasibility queries rather than infeasibility queries. Moreover, \bbdse scales very well, since it does not depend on the trace length but on the user-defined parameter $k$. 
Thus, while backward-bounded DSE seems to be the most appropriate way to solve infeasibility problems
no researches have used this technique.
%
%\forjurnal{Moreover, both techniques could take advantage of each other: DSE better when know infeasibility \cite{icst15}, \bbdse will be more precise if start with more traces}

%\todo[inline]{SB message : ok, c bien pour augmenter coverage / montrer feasibility (orthogonal), et on peut faire preuve d'infeasibility si on enumere tout, mais pas possible en pratique : scale sur un chemin, nb chemins, taint / concret gene correctness of result. ** DSE is done for proving reachability **. Note : DSE and \bbdse complementary : better initial coverage = more paths and less queries. something known about PO or call/ret = less queries to DSE for completion  }

\myparagraph{Backward reasoning}   Backward reasoning is well-known in infinite-state model checking, for example for Petri Nets \cite{FinkelS01}.  \forjournal{WSTS, time automata or pushdown systems}
It is less developed in formal software verification, where forward approaches are prevalent, at the notable exception of deductive verification based on  weakest precondition calculi \cite{Leino05}.    
Interestingly, Charreteur {\it et al.} have  proposed (unbounded) backward symbolic execution  for goal-oriented testing \cite{CharreteurG10}. 
% \forjournal{mcmillan lazy testing}  
% 
Forward and backward approaches are well-known to be complementary, and  can often be combined with benefit \cite{BardinDDKPTM15}.  
%  
%Forward and backward approaches
%
%plus a la mode, except for WP computation (goal-oriented vs forward computation (icst15)). soit focus sur feasibility (BMC, DSE), soit calcul en avant (IA) alors que alternative connue.  
%
% one case of backward DSE, yet: arnaud. but: for feasibility proof. Hence : goal-oriented (good wrt DSE) but  unbounded (no scale) and pure symb (no robust)
%

Yet, purely backward approaches seem nearly impossible to implement at binary level, because of the lack of {\it a priori} information on computed jumps.  
%sur binaire : backward pur impossible because jumps
We solve this problem in \bbdse by performing backward reasoning along some dynamic execution paths observed at runtime, yet at the price of (a low-rate of) false positives.

\myparagraph{Disassembly} Standard disassembly techniques have already been discussed in Section \ref{sec:sparse}.  
Advanced  static techniques include recursive-like approaches extended with patterns dedicated to difficult constructs \cite{binary_not_easy}.   
%static : linear, recursive and combinations, + patterns (ida, objdump, binary code is not easy, etc.) Yet, wellknow to be a big issue on obfuscated code -- VM, po, call/ret, SM, etc. (cf intro / motivation)  -- even instruction overlapping, while easy to handle 
%
Advanced dynamic techniques take advantage of DSE in order to discover more parts of the code \cite{BHKLNPS2007,BardinH11}. 
%
%dynamic : very robust, very safe, but may miss a lot and we do not know where (a branch not taken: a PO or not?)   
%can be clearly enhance by DSE (date back to osmose-journal, bitblaze, etc. ), yet still miss somethings (coverage not enough, infeasibility)
%\bbdse is orthogonal and complementary. infeasibility questions, scales, precise                (new spot)
%
Binary-level semantic program analysis methods \cite{BalakrishnanR10,KinderV10,SeppMS11,BardinHV11,ReinbacherB11} does allow in principle a guaranteed exhaustive disassembly. Even if some interesting case-studies 
have been conducted,    
these methods still face big issues in terms of scaling and robustness. Especially, self-modification is very hard to deal with. The domain is recent, and only very few work exist in that direction \cite{BlazyLP14,giaco}.  
%
%  : ok for medium-size managed code. more guarantees but much more involving. pb with scale and robustness (especially SM code) and precision, especially , but would be great to have formal guarantees. non-obfuscated code already an issue (computed goto). recent work toward SM (laporte) but very preliminary
%
%
%
Several works attempt to combine static analysis and dynamic analysis in order to get better disassembly. Especially, \codisasm\ \cite{codisasm} take advantage 
of the dynamic  trace to perform  syntactic static disassembly of self-modifying programs.

Again, our method is complementary to all these approaches which are mainly based on forward reasoning~\cite{mizuhito}.

%% \myparagraph{Combining static and dynamic disassembly} 
%% Codisasm try to get the best of both world: tracexxxx dynamic analysis to identify SM, then codisasm for disassembly by handling overlapping and SM portions. try to enlarge results of dynamic : yet, big problems with protections such as PO

%% clearly in the vein of our sparse disassemly philosophy. Actually, codisasm could benefit from both DSE (more initial coverage) and \bbdse (sparse)

\myparagraph{Obfuscations} \label{sec:related_obfu}
Opaque predicates were introduced by Collberg~\cite{Collberg1998} giving
a detailed theoretical description and possible usages~\cite{Myles2006, water}
like watermarking. In order to detect them various methods have been proposed~\cite{madou},
notably by abstract interpretation~\cite{giaco} and in recent work with DSE~\cite{loop}. 
Issues raised by  stack tampering and most notably non-returning functions
are discussed by Miller~\cite{binary_not_easy}.  Lakhotia~\cite{lakhotia_call} proposes a method based on abstract interpretation~\cite{lakhotia_call}.  
None of the above solutions address the problem  in such a 
scalable and robust way as \bbdse does.

\section{Conclusion} \label{sec:conclusion} 
Many problems arising during the reverse of obfuscated codes come down to solve infeasibility questions. 
Yet, this class of problem is mostly a blind spot of both standard and advanced disassembly tools. 
% 
%In this article, we have proposed
We propose  
{ Backward-Bounded DSE}, a precise, efficient, robust and generic method for solving infeasibility questions related to deobfuscation. %disassembling  obfuscated codes. %  disassembly.  
We have demonstrated the benefit of the method for several  realistic classes of obfuscations such as  opaque predicate and call stack tampering, and given   insights for other protection schemes. 
%self-modification matters. % the identification of  opaque predicates} (no false  positive, very few false negative) and for the 
%classification of {\tt \small return} statements in case of {\it call stack tampering}. 
%
%
%Finally,  %.  which is   , a combination of backward-bounded DSE and static disassembly able to {\it complete}  dynamic disassembly {\it in a safe way}, hence 
%getting the best of dynamic and static disassembly.   
%
%
%
% 
Backward-Bounded DSE does not supersede existing disassembly approaches, but rather  complements them  by addressing infeasibility questions. % in a scalable and precise manner. 
Following this line, we showed how these techniques can be used to address state-sponsored malware (\xtunnel) and how to merge
the technique with standard static disassembly and dynamic analysis, in order to enlarge a dynamic analysis in a precise and guaranteed  way. % into a  sparse disassembly of heavily obfuscated programs. %, with the best of static and dynamic approaches. 
This work paves the way for precise, efficient and  disassembly tools for obfuscated binaries. 
%

%% The backward-bounded DSE, performed well at detecting the
%% opaque predicates and the various stack tampering technics.
%% All that, scaling on packer size binary programs thanks to
%% the $pre^{\leq k}$ computation. In a future work this algorithm
%% will be extended to target other kind of obfuscations. All that,
%% in order to generate more accurate malware signatures~\cite{codisasm}.
%% \mynote{more.}

\bibliographystyle{IEEEtran}
\bibliography{biblio}

 \clearpage

 \newpage
%\vspace{4em}

 \appendix

%\vspace{-2mm}

\begin{table*}[b!htp]
\caption{Packer experiment: Opaque Predicates \& Call stack tampering}\label{tab:packers}
\centering
\resizebox{1.0\linewidth}{!}{
\def\arraystretch{1.2}
\begin{tabular}{|l|r|r|c|c|c|c|c|c|c|c|c|c|}
\hline
\multirow{3}{*}{Packers} & Static & \multicolumn{5}{c|}{Dynamic} & \multicolumn{6}{c|}{Obfuscation detection} \\
\cline{2-13}
& size        & & & &  & self-mod. & \multicolumn{4}{c|}{Opaque Predicates ($k_{16}$)} & \multicolumn{2}{c|}{Stack tampering} \\
\cline{8-13}
               & prog  &  \#tr.len  & (tr.ok/host)        & \#proc&\#th& \#layers& OK & OP & To & Covered & OK (a/d) & Viol (a/d/s) \\
\hline
ACProtect v2.0    & 101K  & 1.813.598  & (\good,\bad)  & 1 & 1 & 4   & 74 & 159 & 0 & 9   & 0 (0/0) & 48 (45/1/45)\\
Armadillo v3.78   & 460K  & 150.014    & (\bad,\bad)    & 2 & 11& 1   & 1  & 20  & 0 & 1   & 2 (2/0) & 0 (0/0/0)   \\
Aspack v2.12      & 10K   & 377.349    & (\good,\good)& 1 & 1 & 2   & 32 & 24  & 0 & 136 & 11 (7/0)& 6 (1/4/1)   \\
BoxedApp v3.2     & 903K  & /          & \ (\bad,\bad)$^{*}$& 1 & 15& -   & -  & -   & - & -   & -       & -           \\
Crypter v1.12     & 45K   & 1.170.108  & (\good,\bad)  & - & - & 0   & 263& 24  & 0 & 136 & 125 (94/0) & 78 (0/30/32)\\
Enigma v3.1       & 1,1M  & 10.000.000 & \ (\bad,\bad)\sdag    & - & - & 1   & -  & -   & - & -   & -       & -        \\
EP Protector v0.3 & 8,6K  & 250        & (\good,\good)& 1 & 1 & 1   & 10 & 1   & 0 & 2   & 4 (2/0) & 0 (0/0/0)   \\
Expressor         & 13K   & 635.356    & (\good,\good)& 1 & 1 & 1   & 42 & 8   & 0 & 39  & 14 (10/0)& 0 (0/0/0) \\
FSG v2.0          & 3,9K  & 68.987     & (\good,\good)& 1 & 1 & 1   & 11 & 1   & 0 & 14  & 6 (4/0) & 0 (0/0/0) \\
JD Pack v2.0      & 53K   & 42         & (\bad,\good)  & 1 & 1 & 0   & 2  & 0   & 0 & 0   & 0 (0/0) & 0 (0/0/0) \\
Mew               & 2,8K  & 59.320     & (\good,\good)& - & - & 1   & 11 & 1   & 0 & 18  & 6 (4/0) & 1 (0/0/0) \\
MoleBox           & 70K   & 5.288.567  & \ (\good,\good)\sddag& 1 & 1 & 2   & 307& 60  & 0 & 128 & X       & X         \\
Mystic            & 50K   & 4.569.154  & \ (\good,\good)\sddag& 1 & 1 & 1   & X  & X   & X & X   & X       & X         \\
Neolite v2.0      & 14K   & 42.335     & (\good,\good)& 1 & 1 & 1   & 95 & 1   & 0 & 42  & 9 (3/0) & 0 (0/0/0) \\
nPack v1.1.300    & 11K   & 138.231    & (\good,\good)& 1 & 1 & 1   & 41 & 2   & 0 & 34  & 21 (14/0)& 1 (0/0/0)\\
Obsidium v1364    & 116K  & 21         & (\bad,\good)  & - & - & 0   & 1  & 0   & 0 & 0   & 0 (0/0) & 0 (0/0/0) \\
Packman v1.0      & 5,9K  & 130.174    & (\good,\good)& 1 & 1 & 1   & 12 & 1   & 0 & 21  & 7 (4/0) & 0 (0/0/0) \\
PE Compact v2.20  & 7,0K  & 202        & (\good,\good)& 1 & 1 & 1   & 11 & 1   & 0 & 1   & 4 (2/0) & 0 (0/0/0) \\
PE Lock           & 21K   & 2.389.260  & (\good,\good)& 1 & 1 & 6   & 53 & 90  & 0 & 42  & 4 (3/0) & 3 (0/1/0) \\
PE Spin v1.1      & 26K   & /          & \ (\bad,\bad)$^{*}$& 1 & 1 & -   & -  & -   & - & -   & -       & -         \\
Petite v2.2       & 12K   & 260.025    & (\bad,\bad)    & 1 & 1 & 0   & 60 & 19  & 0 & 45  & 4 (1/0) & 0 (0/0/0) \\
RLPack            & 6,4K  & 941.291    & (\good,\good)& 1 & 1 & 1   & 21 & 2   & 0 & 25  & 14 (8/0)& 0 (0/0/0) \\
Setisoft v2.7.1   & 378K  & 4.040.403  & \ (\bad,\bad)\sddag    & 1 & 5 & 4   & X  & X   & X & X   & X       & X         \\
svk 1.43          & 137K  & 10.000.000 & \ (\bad,\good)\sdag  & - & - & 0   & -  & -   & - & -   & -       & -         \\
TELock v0.51      & 12K   & 406.580    & (\bad,\good)  & 1 & 1 & 5   & 0  & 2   & 0 & 5   & 3 (3/0) & 1 (0/1/0) \\
Themida v1.8      & 1,2M  & 10.000.000 & \ (\bad,\good)\sdag  & 1 & 28& 0   & -  & -   & - & -   & -       & -         \\
Upack v0.39       & 4,1K  & 711.447    & (\good,\good)& 1 & 1 & 2   & 11 & 1   & 0 & 30  & 7 (5/0) & 1 (0/0/0) \\
UPX v2.90         & 5,5K  & 62.091     & (\good,\good)& 1 & 1 & 1   & 11 & 1   & 0 & 26  & 4 (2/0) & 0 (0/0/0) \\
VM Protect v1.50  & 13K   & /          & \ (\bad,\good)$^{*}$& 1 & 1 & 0   & -  & -   & - & -   & -       & -         \\
WinUPack          & 4,0K  & 657.473    & (\good,\good)& 1 & 1 & 2   & 12 & 1   & 0 & 33  & 7 (5/0) & 1 (0/0/0) \\
Yoda's Crypter v1.3    & 12K & 240.900 & (\bad,\good)  & 1 & 1 & 3   & 38 & 1   & 0 & 16  & 4 (3/0) & 9 (0/1/0) \\
Yoda's Protector v1.02 & 18K & 17      & (\bad,\good)  & 1 & 1 & 0   & 1  & 0   & 0 & 0   & 0 (0/0) & 0 (0/0/0) \\ 
\hline
\end{tabular}
}

\medskip
\begin{itemize}
	\item \textbf{size prog}: size of the program %\qquad -- \qquad \textbf{\#tr.len}: execution trace length 
	\item \textbf{\#tr.len}: execution trace length 
	\item \textbf{tr.ok}: whether the executed trace was successfully gathered without exception/detection
	\item \textbf{host}: whether the payload was successfully executed \textit{(printing the hostname of the machine)}
	\item \textbf{\#proc}: number of process spawned  %\qquad -- \qquad  \textbf{\#th}: number of threads spawned    \qquad -- \qquad \textbf{\#layers}: number of self-modification layers recorded
	\item \textbf{\#th}: number of threads spawned
	\item \textbf{\#layers}: number of self-modification layers recorded

	\item \textbf{OK, OP, To, Covered}: predicate ok, opaque predicate, timeout, predicate fully covered (both branches)
	\item \textbf{(a/d/s)}: (aligned/disaligned/single)
	\item $\bf*$ failed to record the trace  %\qquad -- \qquad \sdag\ maximum trace length reached {\it (thus packer not analyzed)}      \qquad -- \qquad    \sddag\ analysis failed (due to lack of memory)
	\item \sdag\ maximum trace length reached {\it (thus packer not analyzed)}
	\item \sddag\ analysis failed (due to lack of memory)
\end{itemize}

%	\#tr.len: execution trace length -- tr.ok: whether the executed trace was gather without exception/detection --
%	host: whether the payload was successfully executed (printing the hostname of the machine) -- \#proc: number of
%	process spawned -- \#th: number of threads spawned -- \#layers: number of self-modification layers recorded
%	-- (a/d/s): (aligned/disaligned/single)
\end{table*}

\end{document}